# Ion sources for high-power hadron accelerators

*Daniel C. Faircloth*
Rutherford Appleton Laboratory, Chilton, Oxfordshire, UK

**Abstract**
Ion sources are a critical component of all particle accelerators. They create the initial beam that is accelerated by the rest of the machine. This paper will introduce the many methods of creating a beam for high-power hadron accelerators. A brief introduction to some of the relevant concepts of plasma physics and beam formation is given. The different types of ion source used in accelerators today are examined. Positive ion sources for producing $H^+$ ions and multiply charged heavy ions are covered. The physical principles involved with negative ion production are outlined and different types of negative ion sources are described. Cutting edge ion source technology and the techniques used to develop sources for the next generation of accelerators are discussed.

## 1 Introduction

### 1.1 Ion source basics

An ion is an atom or molecule in which the total number of electrons is not equal to the total number of protons, thus giving it a net positive or negative electrical charge. The name ion (from the Greek ιον, meaning "going") was first suggested by William Whewell in 1834. Michael Faraday used the term to refer to the charged particles that carry current in his electrolysis experiments.

Ion sources consist of two parts: a plasma generator and an extraction system.

The plasma generator must be able to provide enough of the correct ions to the extraction system. There are numerous ways of making plasma: electrical discharges in all of their forms; heating by many different means; using lasers; or even being hit by beams of other particles. The key factor is that the plasma must be stable for long enough to extract a beam for whatever the application requires.

The extraction system must be able produce a beam of the correct shape and divergence angle to the next phase of the accelerator by extracting the correct ions from the plasma and removing any unwanted ions, electrons or neutral particles.

There is a large range of different ion sources out there with many varied applications. Some need to produce ions from tiny samples so they can be analysed and measured. Some need to produce ions for industrial processes such as coating, etching or implanting. Some will go into space to provide thrust for satellites and spaceships. Fusion research demands ion sources that generate huge currents of hundreds of amps with beam cross sections measured in square metres. Radioactive rare isotope ion sources for fundamental research need an entire accelerator facility as one of their key components.

This paper will concentrate on sources for high-power hadron particle accelerators and so will focus on high-current, low-emittance sources. All of the sources in this paper can be configured to produce singly charged positive ions (e.g. $H^+$, $D^+$, $Li^+$), some are better suited to produce multiply charged heavy positive ions (e.g. $Pb^{27+}$) and some can create significant quantities of negative ions (e.g. $H^-$). There are lots of hybrid sources that combine features from different types of source. For the sake of simplicity this paper attempts to concentrate only on the archetypal source types.

## 1.2 History

The first low-pressure discharges were produced by Heinrich Geißler and Julius Plücker in Germany in the mid 1850s. Geißler was a glass blower and inventor commissioned by Plücker to make evacuated glass tubes for his experiments on electric discharges at the University of Bonn. Geißler and Plücker invented a mercury displacement pump that could produce previously unattainable low pressures of less than 100 Pa (1 mbar). The tubes, with electrodes at either end, could be filled with different gases and then evacuated. When a current was passed through them a glow discharge was formed. This allowed Plücker to perform the first experiments in plasma physics. He demonstrated that the plasma could be affected by magnetic fields. The ethereally glowing Geißler tubes (as they became known) were popular mid 19th century entertainment devices, but they also opened the door to the experimentalists that would usher in the atomic age.

In 1869 Johann Hittorf spotted cathode rays in a Geißler tube, but it was William Crookes in early 1870s London that first produced them without the glow discharge. Crookes used a modified Geißler tube and an improved mercury pump made by Hermann Sprengel. He was able to obtain pressures as low as 1 Pa ($1 \times 10^{-5}$ mbar). At these very low pressures the glow discharge stops, leaving a pure cathode ray (electron) source.

Thermionic emission of electrons had first been observed by Fredrick Guthrie in 1868. He noticed that red hot metal balls lost charge. Hittorf and various other German researchers also investigated the phenomenon, but it was Thomas Edison that really developed the idea when trying to work out how to improve his light bulbs in 1880.

British and German researchers continued experimenting with vacuum tubes and in 1886 Eugen Goldstein discovered that a perforated cathode (with holes in it) could also emit a beam: he called them anode rays or canal rays (because they emerge from channels in the cathode) and these rays turned out to be positive ions. A schematic of a canal ray ion source is shown in Fig. 1.

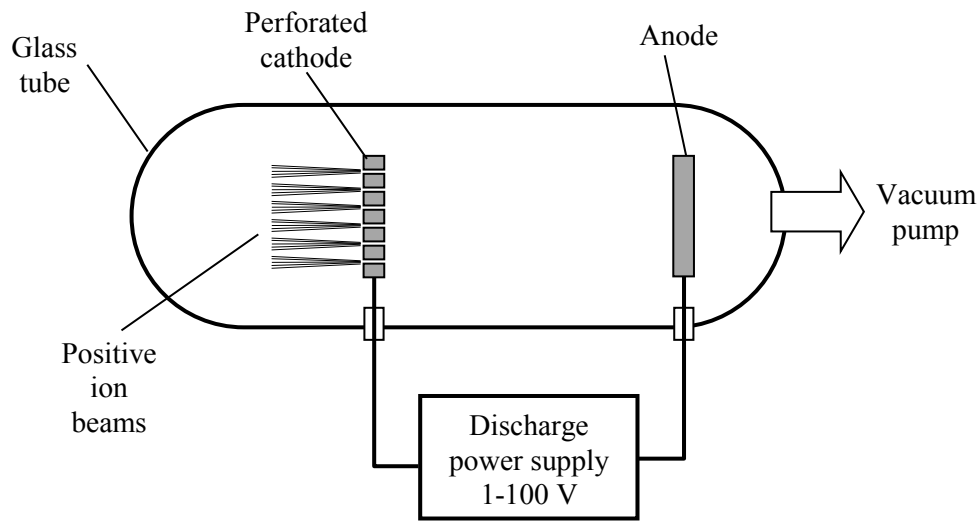

**Fig. 1:** Schematic of a canal ray tube: the first positive ion source

Eventually, in 1897 J.J. Thomson proved that the cathode rays were actually negatively charged particles which were later named electrons. Experiments with magnetic and electrostatic deflection of the newly produced beams of particles led to new theories on the nature of matter. In the early 20th century, a drive to understand the structure within the atom caused researchers to try and further accelerate beams of particles. Different devices and machines were developed and the ion source as we think of it today was born as a means to produce beams of particles.

## 2    Plasma

### 2.1    Introduction

Plasma is the fourth state of matter: if you keep heating a solid, liquid or a gas it will eventually enter the plasma state. Plasma consists of both negatively and positively charged particles in approximately equal proportions, along with un-ionized neutral atoms and molecules. The charged particles consist of positive ions, negative ions and electrons. The particles are always interacting with each other. Collisions can cause ionization or neutralization. Atoms and molecules can be put into excited states and can absorb and emit photons.

The physics of plasmas can be extremely complex. What follows are some of the key concepts relating to ion sources. Plasmas exist in nature wherever the temperature is high enough. Some examples are shown in Fig. 2. The two basic parameters that define a plasma are density and temperature.

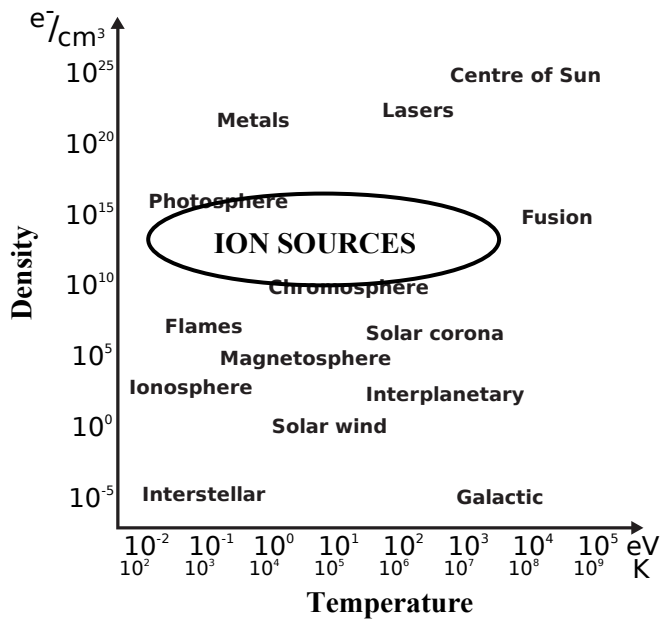

**Fig. 2:** Different types of plasma

### 2.2    Basic plasma parameters

#### 2.2.1    Density, n

The most basic parameter is the density of each of the constituents in the plasma. It is usually written as $n$ with subscript to represent the type of particle and is expressed in number of particles per cubic meter. Some older papers give density in particles per cubic centimeter.

$n_e$ = density of electrons
$n_i$ = density of ions
$n_n$ = density of neutrals

#### 2.2.2    Temperature, T

The temperature of the plasma is a measurement of how fast each of the particles is going, otherwise known as the particle kinetic energy. The Boltzmann constant gives 11 600 K = 1 eV. The temperature is usually expressed in electronvolts:

$T_e$ = temperature of electrons
$T_i$ = temperature of ions
$T_n$ = temperature of neutrals

The temperatures of the electrons, ions and neutrals can be different. Different types of plasma produced in different ion sources can have very different ion and electron temperatures. For example, electron cyclotron resonance (ECR) ion sources (see Section 4.4.2), where the plasma is heated by accelerating the electrons, can have $T_e > 1$ keV and $T_i < 1$ eV.

### 2.2.3    *Charge state, q*

The charge state of the ions is also important when defining the properties of a plasma. The charge state of an ion indicates how many electrons have been removed from it. Singly charged ions have a charge state $q = +1$. Not all ions will be singly charged, some ions will be multiply ionized (e.g. $Pb^{3+}$ which has a charge state $q = +3$). Some ions will be negatively charged (e.g. $H^-$ which has a charge state $q = -1$). The densities, $n$, of each charge state can be very different.

Some sources are designed to produce beams of ions with very high charge state, such as $Ag^{32+}$ ions from an electron beam ion source (EBIS; see Section 4.5).

## 2.3    Ionization energy

The ionization energy is the energy in electronvolts required to remove an electron from an atom. The larger the atom, the easier it is to remove the outermost electron. The second electron is always harder to remove, the third even harder and so on. In most ion sources the energy for ionization comes from electrons impacting on neutral atoms or molecules in electrical discharges. The electrons receive their energy from being accelerated by the field applied to the discharge.

## 2.4    Temperature distributions

Obviously not all electrons in plasma will have the same temperature and the same is true for the ions of the same species. The numbers $T_e$, $T_i$ and $T_n$ are merely averages. If the plasma is in thermal equilibrium then the distribution will be Maxwellian and obey Maxwell–Boltzmann statistics.

Using the standard equations the mean speeds of the ions can be calculated to be:

$$\text{velocity of electrons, } \bar{v}_e = 67\sqrt{T_e} \tag{1}$$

$$\text{velocity of ions, } \bar{v}_i = 1.57\sqrt{\frac{T_i}{A}} \tag{2}$$

where $A$ is the ion mass in atomic mass units.

Often the plasma is in a magnetic field. The ions and electrons will spiral around the magnetic field lines and slowly move along them. Hence, the particle velocities (temperatures) will not be the same in all directions. The particle temperatures are defined as $T_{i\parallel}$ is the ion temperature parallel to the magnetic field and $T_{i\perp}$ is the ion temperature perpendicular to the magnetic field.

## 2.5    Quasi-neutrality

Plasma is generally charge neutral, so all of the charge states of all the ions adds up to the same number as the number of electrons:

$$\sum q_i \, n_i = n_e \tag{3}$$

## 2.6 Percentage ionization

The percentage ionization is a measure of how ionized the gas is, i.e. what proportion of the atoms have actually been ionized:

$$\text{percentage ionization} = 100 \times \frac{n_i}{n_i + n_n} \tag{4}$$

When the percentage is above 10% the plasma is said to be highly ionised and the interactions that take place within are dominated by plasma physics. Less than 1% ionisation and interactions with neutrals must be considered.

## 2.7 Electrical discharges

### 2.7.1 Overview

The driving field applied to a discharge is often the electrical field; however the magnetic field can also be used in the case of inductively coupled discharges. Inductively coupled discharges require a time varying field and so are more difficult to analyse. It is best to start with an explanation of a DC electric-field-driven discharge between two electrodes. The general current and voltage characteristics of such a discharge are summarized in Fig. 3. The exact shape of the curve depends on the type of gas, pressure, electrode geometry, electrode temperatures, electrode materials, and any magnetic fields present.

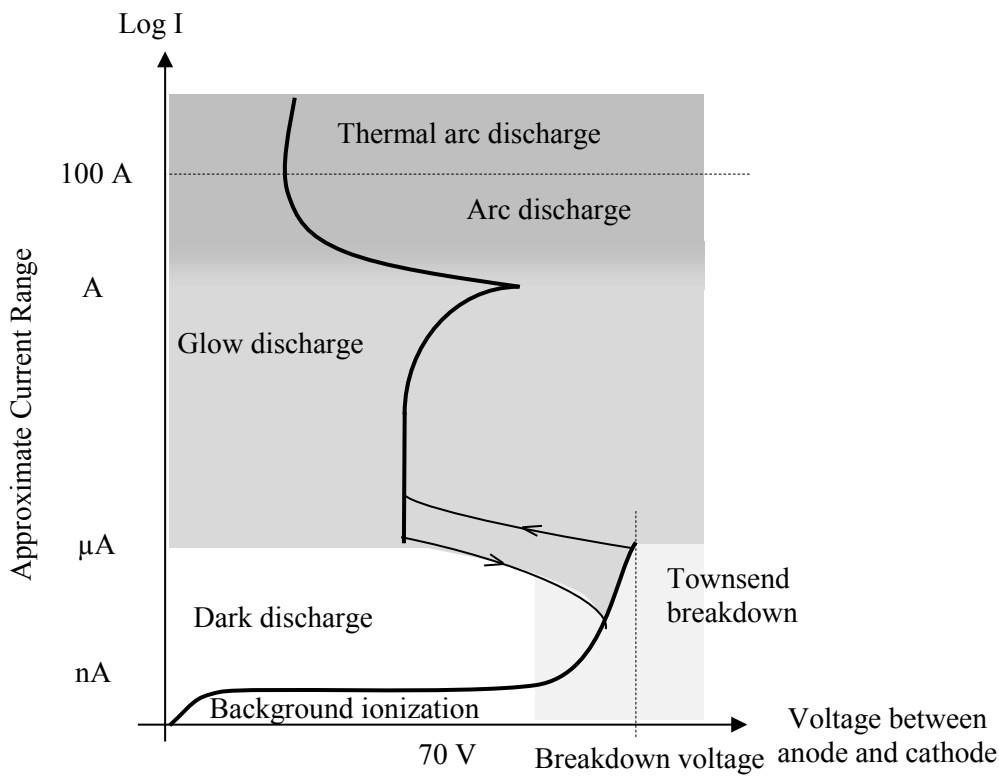

**Fig. 3:** The current voltage characteristics of a typical electrical discharge

### 2.7.2 Dark discharge

At low voltages the current between two electrodes is very small, but it slowly increases as the voltage between the electrodes increases as shown in the bottom left corner of Fig. 3. This tiny current comes from ions and electrons produced by background ionization. These are swept out of the gap by the electric field between the electrodes that is created by the applied voltage. There are only enough charge carriers produced by background radiation for a few nanoamps of current, so the current quickly saturates. The voltage can then be increased with no increase in current. The ions and electrons are pulled towards the electrodes through the gas molecules interacting with them as they go.

### 2.7.3 Townsend breakdown

Eventually the applied electric field is high enough to accelerate the electrons to the ionization energy of the gas. At this point the current rapidly increases as shown in bottom right corner of Fig. 3. The electrons ionize the neutral atoms and molecules, producing more electrons. These additional electrons are accelerated to ionize even more atoms producing even more free electrons in an avalanche breakdown process known as Townsend breakdown. This runaway process means the voltage needed to sustain the discharge drops significantly. The discharge has entered the glow discharge regime.

### 2.7.4 Glow discharge

The glow discharge is so called because it emits a significant amount of light. Most of the photons that make up this light are produced when atoms that have had their orbital electrons excited by electron bombardment, relax back to their ground states. Photons are produced in any event that needs to release energy, for example when ions recombine with the free electrons and when vibrationally excited molecules relax.

A glow discharge is self-sustaining because positive ions that are accelerated to the cathode impact, producing more electrons in a process called secondary emission. This is why there is a hysteresis in the current versus voltage curve at the glow-to-dark discharge transition.

The current in a glow discharge can be increased with very little increase in discharge voltage. The plasma distributes itself around the cathode surface as the current increases. Eventually the current reaches a point where the cathode surface is completely covered with plasma and the only way to increase the current further is to increase the current density at the cathode. This causes the plasma voltage near the cathode to rise.

### 2.7.5 Arc discharge

The increased current density leads to cathode heating and eventually the cathode surface reaches a temperature where it starts to thermionically emit electrons and the discharge moves into the arc regime with a negative current versus voltage characteristic.

The current increases until plasma is almost completely ionized (there are few neutral particles left). Eventually the current density in the plasma reaches a point where the ions have the same average velocity as the electrons: they have reached thermal equilibrium. The discharge enters the thermal arc regime where the discharge voltage rises as the current increases.

### 2.7.6 Importance of the power supply

The power supply used to produce the discharge will have a large effect on the type of discharge produced. The discharge current and voltage obtained will be where the power supply load curve intersects the characteristic shown in Fig. 3. The gradient of the discharge characteristic at the intersection point determines whether the discharge is stable or not. Most ion sources operate in the glow regime.

## 2.8    Paschen curve

The breakdown voltage of any gas between two flat electrodes depends only on the electron mean free path and the distance between the electrodes. The mean free path is the average distance particles travel before hitting other particles. It is directly related to pressure. Figure 4 shows how the breakdown voltage of hydrogen varies with the product of pressure, $p$, and distance, $d$, between electrodes. This was first stated in 1889 by Friedrich Paschen [1].

At very low pressures, the mean free path between collisions is longer than the distance between the electrodes. So although the electrons can be accelerated to ionising energies, they are unlikely to hit anything other than the anode. This means that the breakdown voltage is very high at very low pressures.

At very high pressures the mean free path is very short. This means that the electrons never have enough time to be accelerated before hitting another particle. This means at the breakdown voltage is high at very high pressures.

Between these two extremes is a minimum whose position depends on the type of gas and the electrode material. This "Paschen minimum" leads to a counterintuitive phenomenon: operating just below this minimum, electrodes further apart will have a lower breakdown voltage than those closer together. This is because in a longer gap there is more space for the electron avalanches to develop.

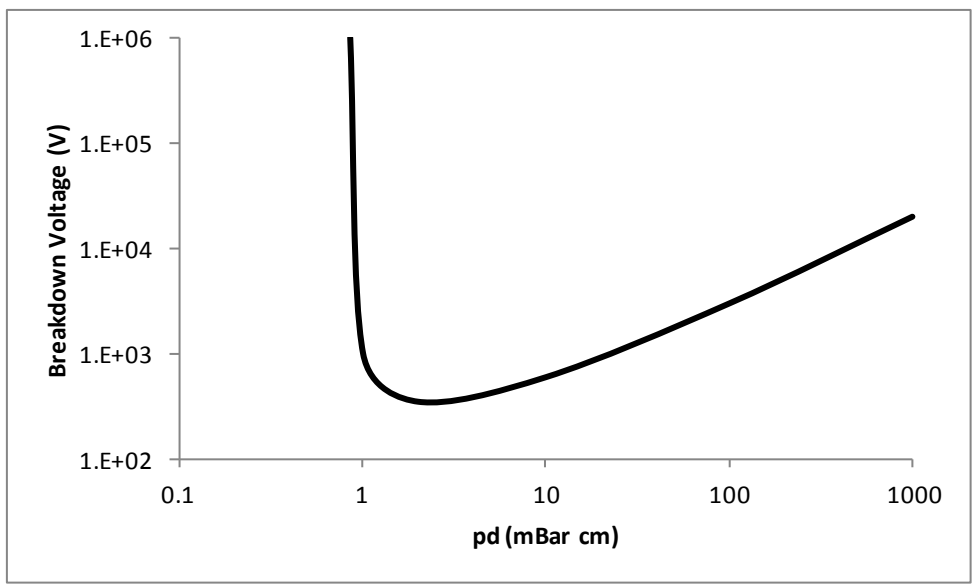

**Fig. 4:** The Paschen curve for hydrogen

## 2.9    Collisions

Collisions between particles in a plasma are fundamentally different from collisions in a neutral gas. The ions in a plasma interact by the Coulomb force: they can be attracted or repelled from a great distance. As an ion moves in a plasma its direction is gradually changed as it passes the electric fields of neighbouring particles, whereas in a neutral gas the particles only interact when they get so close to each other that they literally bounce off each other's outer electron orbitals. In a neutral gas the average distance the particles travel in a straight line before bouncing off another particle is referred to as the mean free path. In a plasma, the mean free path concept does not work because the ions and electrons are always interacting with each other by their electric fields. Instead a concept called "relaxation time" is invoked: this is the time it takes for an ion to change direction by 90°. The relaxation time $\tau_0$ can also be described as "the average 90° deflection time". In a plasma there are different relaxation times between each of the different particle species.

## 2.10 Work function

In any solid metal, there are one or two electrons per atom that are free to move from atom to atom. This is sometimes collectively referred to as a "sea of electrons". Their velocities follow a statistical distribution, rather than being uniform. Occasionally an electron will have enough velocity to exit the metal without being pulled back in. The minimum amount of energy needed for an electron to leave a surface is called the work function. Specifically the work function is the energy needed to move an electron from the Fermi level into vacuum. The work function is characteristic of the material and for most metals is of the order of several electronvolts.

## 2.11 Thermionic emission

Thermionic emission is the heat-induced flow of charge carriers from a surface or over a potential-energy barrier. This occurs because the thermal energy given to the carrier overcomes the work function of the metal. Thermionic currents can be increased by decreasing the work function. This often-desired goal can be achieved by applying various oxide coatings to the wire.

In 1901 Owen Richardson found that the current from a heated wire varied exponentially with temperature. He later proposed this equation:

$$J = A_G T^2 e^{\frac{-W}{kT}} \tag{5}$$

where $J$ is the electron current density on the surface of the cathode, $W$ is the cathode work function and $T$ is the temperature of the cathode.

Here $A_G$ is given by

$$A_G = \lambda_R A_0 \tag{6}$$

where $\lambda_R$ is a material-specific correction factor that is typically of order 0.5 and $A_0$ is a universal constant given by

$$A_0 = \frac{4\pi m k^2 e}{h^3} = 1.20173 \times 10^6 \ \text{Am}^{-2}\text{K}^{-2} \tag{7}$$

where $m$ and $e$ are the mass and charge of an electron and $h$ is Planck's constant.

## 2.12 Magnetic confinement

Charged particles will rotate around magnetic field lines: this means that they tend to travel along magnetic field lines, by spiralling along them. This effect can be exploited to confine plasma in an ion source. A dipole field will confine particles in the direction of the magnetic field. This can be used to confine electrons between two parallel cathodes, as employed in the Penning ion source (Section 5.3.3). A solenoid field will keep charged particles confined axially. Solenoidal fields are used in duoplasmatrons (Section 4.3.2), microwave ion sources (Section 4.4), EBISs (Section 4.5) and vacuum arc ion sources (Section 4.7).

A multicusp field is composed of alternating north and south poles (see Section 5.4). This arrangement is used around the edge of plasma chamber to confine both electrons and ions and prevents them from hitting the walls of the chamber. A specific type of multicusp field (the hexapole field) is used in to increase the confinement time in ECR ion sources (Section 4.4.2).

## 2.13 Debye length

Named after the Dutch scientist Peter Debye, the Debye length, $\lambda_D$, is the distance over which the free electrons redistribute themselves to screen out electric fields in plasma. This screening process occurs because the light mobile electrons are repelled from each other whilst being pulled by neighbouring heavy low-mobility positive ions, thus the electrons will always distribute themselves between the ions. Their electric fields counteract the fields of the ions creating a screening effect. The Debye length not only limits the influential range that particles' electric fields have on each other but it also limits how far electric fields produced by voltages applied to electrodes can penetrate into the plasma. The Debye length effect is what makes the plasma quasi-neutral over long distances.

The higher the electron density the more effective the screening, thus the shorter this screening (Debye) length will be.

The Debye length is given by

$$\lambda_D = \sqrt{\frac{\epsilon_0 k T_e}{n_e q_e^2}} \tag{8}$$

Where:

$\lambda_D$ is the Debye length,

$\epsilon_0$ is the permittivity of free space,

$k$ is the Boltzmann constant,

$q_e$ is the charge of an electron,

$T_e$ is the temperatures of the electrons

$n_e$ is the density of electrons

$\lambda_D$ is of the order $0.1 - 1$ mm for ion source plasmas.

## 2.14  Plasma sheath

The screening effect of the plasma creates a phenomenon called the plasma sheath around the cathode electrode. The plasma sheath is also called the Debye sheath. The sheath has a greater density of positive ions, and hence an overall excess positive charge. It balances an opposite negative charge on the cathode with which it is in contact. The plasma sheath is several Debye lengths thick.

The quasi-uniform plasma potential is closest to the anode voltage and the largest potential drop in a plasma is across the plasma sheath near the cathode.

A related phenomenon is the double sheath. This occurs when a current flows in the plasma.

## 2.15  Particle feed methods

A supply of material to be ionized must be provided to the plasma. If the material is a gas it can be introduced via a needle valve or mass flow controller. If the source is pulsed the gas is usually also pulsed to help maintain low pressures in the source, which is usually achieved with a fast piezo-electric valve. Some gasses are very corrosive so compounds of the element are used instead.

Some materials can be heated in ovens (e.g. caesium, see Section 5.2.6). Other solid materials with low vapour pressure are more suited to ionization by a laser (Section 4.6) or in an arc discharge (Section 4.7). Some sources (such as the EBIS, see Section 4.5) are often fed by another ion source.

It is important to prevent unionized material and excess ions entering the next stage of the accelerator, this is achieved by having a high pumping speed and vessel constrictions with baffles and cold traps.

# 3    Extraction

## 3.1    Introduction

The purpose of the extraction system is to produce a beam from the plasma generator and deliver it to the next acceleration stage. The basics of extraction are very simple: apply a high voltage between an ion emitting surface and an extraction electrode with a hole in it. The extraction electrode can also be called the acceleration electrode or the ground electrode (if the plasma is produced in a discharge on a high-voltage platform).

## 3.2    Meniscus emitting surface

In plasma sources the ion emitting surface is the edge of the plasma itself. At the extraction region the plasma is bounded by an electrode with a hole in it. This electrode is variously called the outlet electrode, aperture electrode or plasma electrode and it is often at the same potential as the plasma anode. The edge of the plasma sits across this hole and is called the plasma meniscus. It is the boundary layer between the discharge and the beam. The shape of the meniscus depends on the local electric field and the local plasma densities. Figure 5 shows how the plasma meniscus can change from being convex to concave as the plasma density decreases. The trajectories of the particles depend on the meniscus shape, so it is important to run the ion source with operating conditions that provide an optimum meniscus shape. This is called the "matched case" and is found by varying the plasma density and extraction potential until the beam is well transported. It is important to point out that the diagrams in Fig. 5 do not include space charge effects that cause the beam to diverge (see Section 3.7).

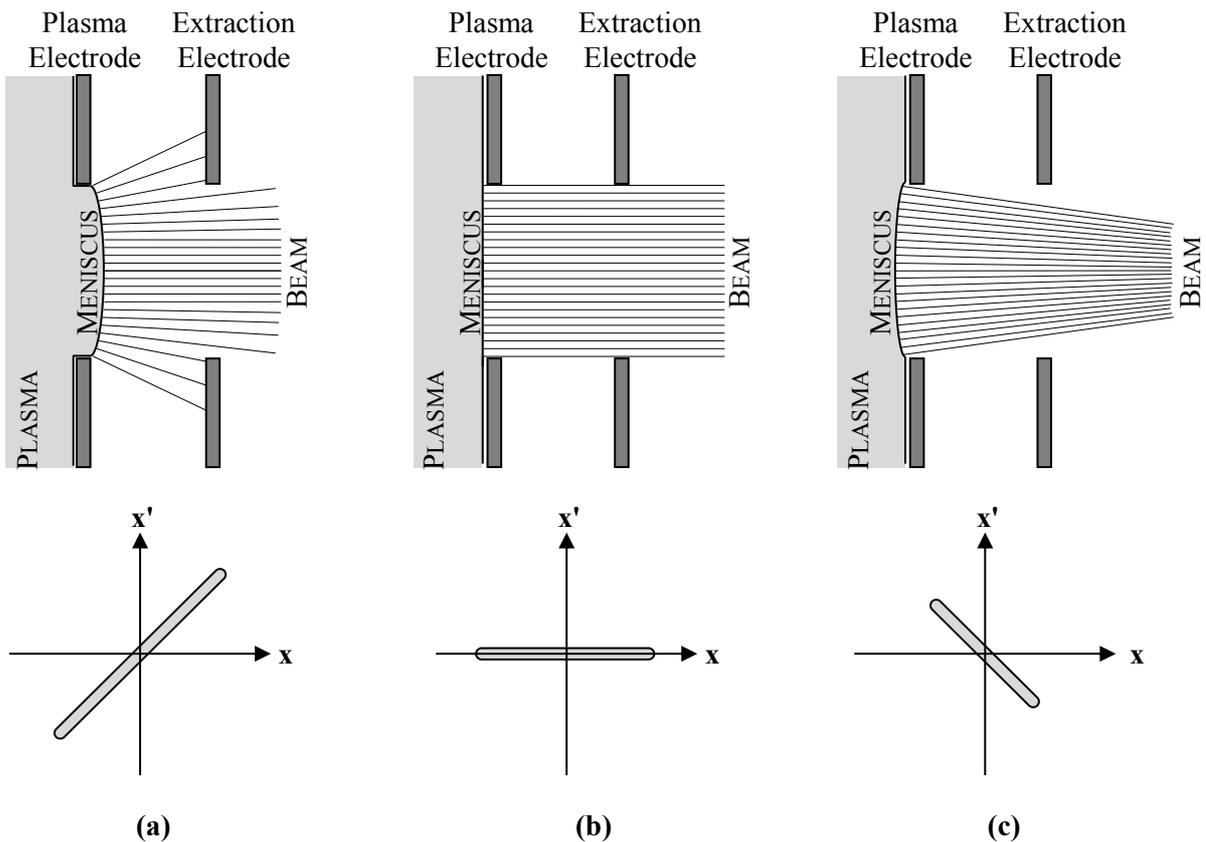

**Fig. 5:** Three different plasma meniscus shapes and their corresponding emittance phases space plots: (a) convex, high plasma density; (b) flat, medium plasma density; (c) concave, low plasma density

### 3.3    Solid emitting surface

In surface converter sources the ions of interest are actually produced on a solid surface inside a plasma chamber (see Section 5.5). This has a great advantage over meniscus emission in that the exact shape of the surface can be precisely defined. Solid emission surfaces are concave with a radius of curvature approximately centred on the exit aperture in the plasma chamber. This causes the ions produced on the emission surface to be focused at the exit aperture, resulting in a high-quality, low-divergence beam.

### 3.4    Emittance

For high-power particle accelerators it is essential that the beam produced by the ion source has a low divergence angle. This allows the beam to be transported and accelerated easily by the rest of the machine without losing any beam.

In particle accelerators a way of specifying the divergence of a beam is emittance. Emittance is a measurement of how large a beam is and how much it is diverging. It is measured in millimetre-milliradians and is the product of beam size and divergence angle. Often emittance is normalized to beam energy because a beam that has had its longitudinal velocity increased by acceleration will still have the same transverse velocity, thus its divergence angle will be reduced. Emittance is normalized to allow emittances to be compared at different energies in different accelerators.

For any beam, two emittances are given: horizontal and vertical. These can be just single values or complete phase space diagrams. Phase space diagrams are plots of divergence angle versus transverse position. Figure 5 gives examples of phase space plots for divergent, parallel and convergent beams. If the horizontal axis of the emittance phase space diagram is $x$ then the vertical axis is usually given as $x'$ (where $x'$ is $\frac{dx}{dz}$ or $\frac{v_x}{v_z}$), this is because the ratio of transverse and longitudinal velocities of the particles defines the divergence angle. The units of $x'$ are radians and are usually expressed in milliradians.

The emittance of a beam is the area enclosed by its phase space plot divided by $\pi$. For real beams this statement needs clarification. Real beams have halos: outlying particles that have much larger divergence angles and positions than the core of the beam. They are created when some particles at the edge of the beam experience fringe fields or other non-uniformarities that cause them to separate further from the core of the beam. The halo particles are in the minority; most of the particles are in the dense core of the beam. If these outlying particles are included in the total phase space area calculation the beam will have a huge emittance. To get round this emittances are either quoted as the 95% emittance or root mean square (r.m.s.) emittance. The 95% emittance is the area that encloses 95% of all of the particles. The r.m.s. emittance is calculated from the r.m.s. values of all of the positions and angle measurements. Both methods give a realistic measurement of the overall beam divergence without being unfairly enhanced by the halo particles.

The unit of emittance is actually distance. This makes sense because the dimensions of phase space are millimetres and a dimensionless angle, hence areas in phase space have units of mm. Ion source emittances, however, are very often expressed in units of $\pi\cdot$mm$\cdot$mrad. This is because the particle distributions in phase space plots often have an ellipse drawn round them to define the beam boundary. The area of an ellipse is $\pi$ multiplied by the product of the length of its two semi-axes. Later, in the rest of the accelerator, higher-energy beams usually do tend to have elliptical phase space distributions. Close to the ion source large aberrations still exist and the beam shape in phase space is often far from elliptical, so a r.m.s integration is the best method to calculate the emittance. It is counterintuitive that r.m.s. ion source emittances are expressed in $\pi\cdot$mm$\cdot$mrad (which is indicative of an ellipse calculation). Units of mm or $\pi\cdot$mm$\cdot$mrad do not change the emittance value quoted and can be used interchangeably. Care must be taken, however, when comparing emittances from older papers that use mm$\cdot$mrad, in this case the emittance is actually $\pi$ times larger than if quoted in mm.

## 3.5 Energy spread

In a beam not all of the particles have the same energy. The energy distribution of the particles is a measurement of the range of different particle velocities in the beam. It is effectively the longitudinal emittance of the beam measured in electronvolts. It is more commonly defined as the full-width–half-maximum of the energy distribution in electronvolts. It is sometimes also referred to as the momentum spread. In emittance phase space plots a proportion of the "thickness" of the phase space distribution is caused by energy spread.

The energy spread causes the beam to spread out in time, so this limits the minimum pulse length achievable. The origin and size of the energy spread is different for different types of source. It can be caused by variations in potentials or temperatures on the plasma production surface, oscillations in the plasma or unstable extraction voltages. The energy spread can range from less than 1 eV to as much as 100 eV.

Energy spread is important because it will produce transverse emittance growth as the beam passes though magnets and accelerating gaps. The beam emittance can be transferred between longitudinal and vertical directions and vice versa.

## 3.6 Brightness

The brightness of a beam is another key beam parameter. It is the beam current, $I$, divided by the emittances:

$$B = \frac{I}{\varepsilon_x \varepsilon_y} \tag{9}$$

Unfortunately there are several ways to define brightness: they all have the same basic form as Eq. (9) but they have each have different scale factors based on multiples of $\pi$ and 2. The reader should take caution when comparing brightness from different authors. Also sometimes the emittances are normalized to energy, which gives an emittance-normalized brightness.

## 3.7 Space charge

Space charge effects are critical in ion source design. For high-brightness, low-energy beams electrostatic forces are a key factor. The beam will blow-up under its own space charge so it is critical to get the beam energy up to at least 10 keV as fast as possible to minimize the effect which is worse at low energies.

A phenomenon known as "space charge compensation" or "space charge neutralization" is essential for high current ion sources. The pressure in the vacuum vessel directly after extraction will be higher than in the rest of the accelerator because of gas loading from the ion source itself. The beam ionizes the background gas as it passes through it. If the beam is positive it repels the positive background ions and draws in the negative ions and electrons. If the beam is negative it repels the negative background ions and electrons and draws in the positive ions. The effect is to neutralize the beam, reducing its space charge and reducing the beam blow-up. Beams can be almost 100% space charge compensated, meaning they see almost no beam blow-up.

Space charge compensation is a complex dynamic process, for pulsed beams it can take about 100 μs to build up the compensation particles so the start of a pulsed beam will have a transient change in emittance. For very long beam pulses (> 1 ms) the beam can actually lose its compensating particles by diffusion a process known as decompensation.

Space charge compensation is not possible in accelerating gaps because any compensating particles produced are swept out of the gap by the field. For the short time compensating particles remain in the gap they are not effective at compensating the beam because they are moving too fast.

### 3.8 Child–Langmuir law

There is an absolute limit to the current density that can be extracted from a plasma. There comes a point where the space charge of the beam being extracted actually cancels out the extraction field, making it impossible to extract a higher current density. The current density where this happens can be calculated from the Child–Langmuir equation:

$$j = \frac{\frac{4}{9}\epsilon_0 \sqrt{\frac{2q_i}{m_i}} V^{\frac{3}{2}}}{d^2} \tag{10}$$

where $j$ is the current density in A m$^{-2}$, $q_i$ is the ion charge in coulombs, $d$ is the extraction gap in metres, $m_i$ is the ion mass in kg and $V$ is extraction voltage in V.

If more useful units are used, Eq. (10) becomes

$$j = \frac{1.72 \sqrt{\frac{q}{A}} V^{\frac{3}{2}}}{d^2} \tag{11}$$

where $q$ is the ion charge state, $A$ is the ion mass in atomic mass units, $d$ is the extraction gap width in cm, $V$ is the extraction voltage in kV and $j$ the current density is now in mA cm$^{-2}$.

These equations are true for space charge limited conditions, i.e. where the plasma generator has plenty of ions to give, but space charge limits the current. If the plasma cannot give any more ions then the source is no longer space charge limited and the current versus voltage relationship shown in Eq. (11) no longer follows.

### 3.9 Perveance

The perveance, $P$, of an ion source is a measurement of how space charge limited the source is. It is defined as

$$P = \frac{I}{V^{\frac{3}{2}}} \tag{12}$$

where $I$ is the beam current.

It is the constant of proportionality in Eq. (12), it should be constant as the extraction voltage is increased. The voltage where $P$ starts to decrease is an indication that the plasma can no longer supply enough particles to the extractor. The word perveance comes from the Latin "pervenio" meaning to attain. Perveance is also called "puissance" in some texts; this is actually a better word as it is French for strength or ability, and perveance refers to the strength or ability of the plasma to deliver ions.

### 3.10 Pierce extraction

The shape of the electric field in the extraction gap will shape the beam as it is extracted. The Pierce electrode geometry is an attempt to produce an absolutely parallel beam. The idea is to produce an extraction field that has a zero transverse value at the edge of the beam, thus not having any focusing effect on the beam. The standard Pierce geometry consists of a plasma electrode at an angle of 67.5° to the beam axis. The extraction electrode is curved along an equipotential line to the solution of the equation that gives zero traverse field at the beam edge.

In reality, a completely parallel beam is impossible.

### 3.11 Suppressor electrode

In accelerating gaps for positive ions, the electrons will be accelerated in the opposite direction and into the ion source. This is not desirable so an electron suppressor electrode is often added just before the ground electrode. The suppressor electrode is biased slightly more negative than the ground electrode. Any electrons heading into the acceleration gap from the ground electrode side will be reflected back as shown in Fig. 6.

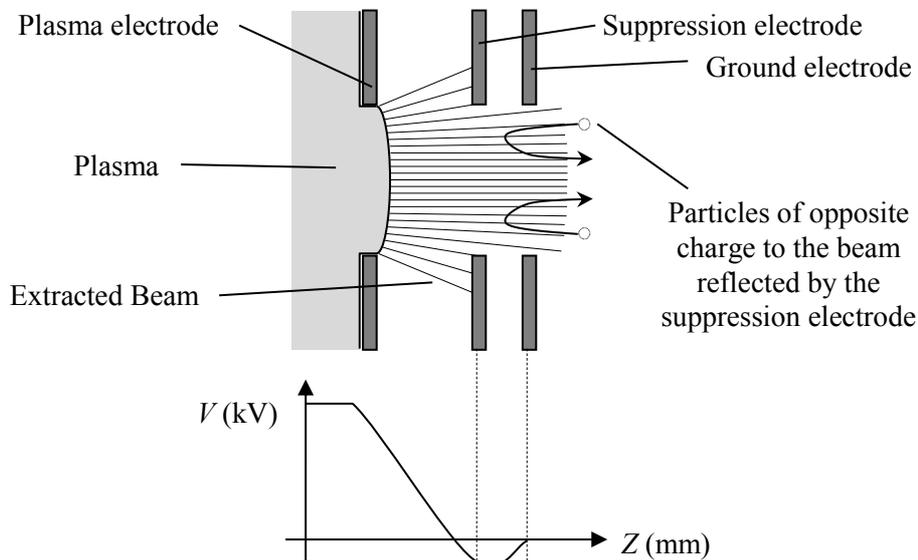

**Fig. 6:** The use of a suppressor electrode to prevent back-streaming particles of the opposite charge from entering the ion source

In negative ion sources, protons will be accelerated back into the source instead of electrons so the suppressor is biased with a positive voltage. Back-streaming particles can damage the source by sputtering so it is important to suppress them.

### 3.12  Negative ion extraction

One of the main challenges with negative ion source design is how to deal with the co-extracted electrons. Extracting electrons with the negative ions is obviously unavoidable because they both have the same charge. The ion source engineer must first try to minimize the amount of co-extracted electrons, then find a way to separate and dump the unwanted electrons from the negative ion beam. In some cases the electron current can be 1000 times greater than the negative ion current itself.

### 3.13  Low-energy beam transport

It could be argued that the ion source extraction system should include the low-energy beam transport (LEBT) system as well. Beam halo and emittance effects mean that the current measured directly after extraction is not a true measure of the beam current that can be transported to the next stage of the accelerator. Often there is significant collimation of the beam on the way through the LEBT and a large proportion of beam current can be lost.

The whole ion source usually sits on a high-voltage platform. The beam is accelerated to ground (this is why the last electrode of the extraction system in Fig. 6 is labelled as the ground electrode), then the beam enters the LEBT. LEBTs can be magnetic or electrostatic or a combination of both. It is common to use between one and four focusing elements. These can be electrostatic Einzel lenses or electromagnetic solenoids and quadrupoles. In ion sources that use caesium vapour it is better to use an electromagnetic LEBT to prevent sparking.

## 4 Positive ion sources

### 4.1 Introduction

Researchers had been experimenting with beams of positive ions (or canal rays as they called them) since 1886 when Eugen Goldstein discovered that they were emitted from holes in the cathode. The problem with canal ray sources was that the beam energy could not easily be varied and they had a huge energy spread- as large as the discharge voltage.

### 4.2 Electron bombardment sources

The first real positive ion source was developed by Arthur Dempster at the University of Chicago in 1916 [2]. The basic design is shown in Fig. 7. It is the first source to introduce an extraction electrode.

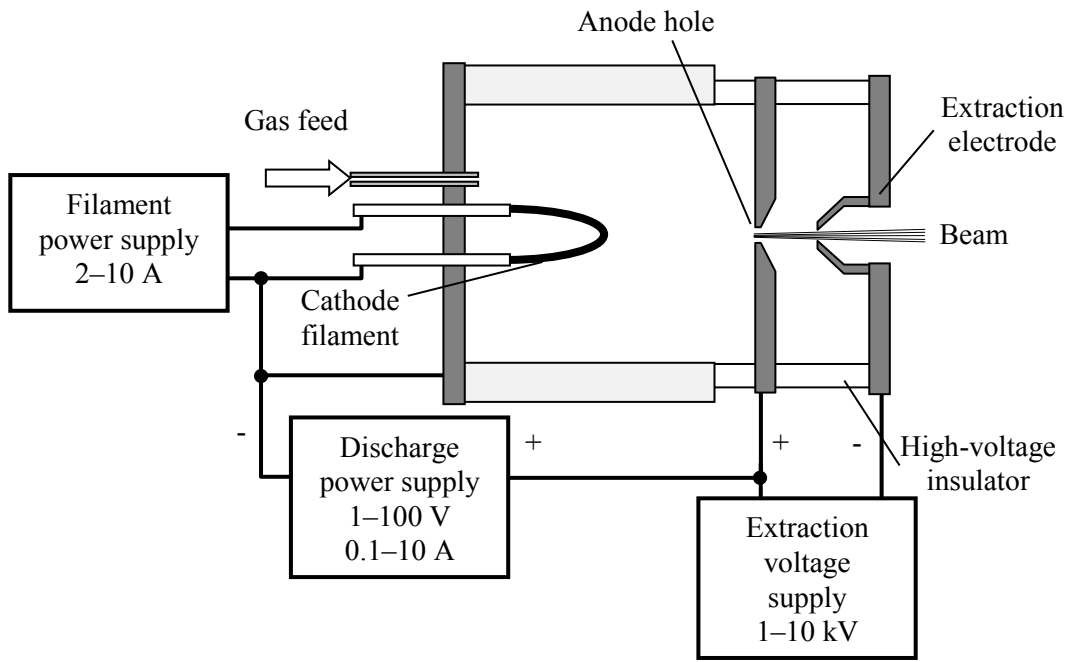

**Fig. 7:** Schematic of an electron bombardment source

The cathode is heated by passing a current so that it thermionically emits electrons. The electrons are accelerated by the discharge power supply voltage. As long the discharge voltage is higher than the ionization energy of the gas fed into the source, the electrons will be able to ionize the gas by impact ionization. The anode has a small hole in it, opposite which there is an extraction electrode. A negative high voltage is applied to the extraction electrode. The positive ions produced near the anode hole are extracted from the source. Beam currents of about 1 mA can be produced.

Electron bombardment sources are very cheap and easy to produce, they can be used to produce positive beams from almost every element, but they cannot generate beam currents of the magnitude required for high-power accelerators.

### 4.3 Plasmatrons

#### 4.3.1 Introduction

The plasmatron was first developed by the prolific aristocratic German inventor Manfred von Ardenne in the late 1940s. It is a development of the electron bombardment source. To increase the beam current, a conical shaped intermediate electrode is positioned between a heated filament cathode and an anode as shown in Fig. 8. The purpose of the conical intermediate electrode is to "funnel" the plasma down to a higher-density region near the anode extraction hole. A plasma double sheath forms on the conical intermediate electrode. The higher plasma density near the extraction region allows more ions to be extracted.

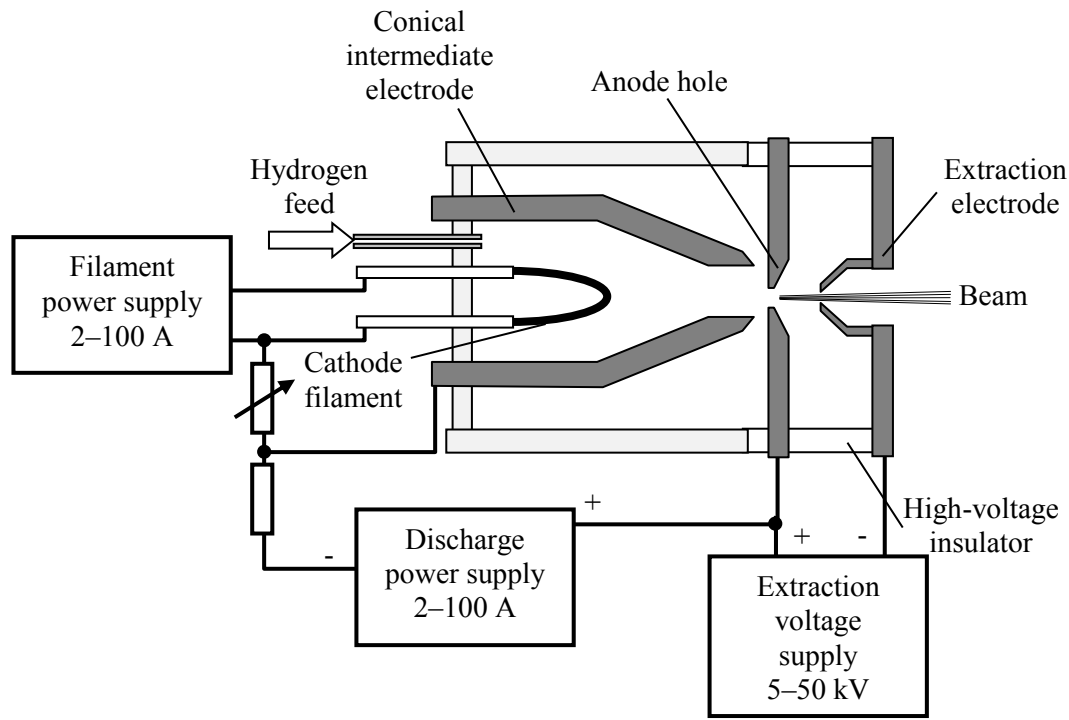

**Fig. 8:** A schematic of a plasmatron source

### 4.3.2    Duoplasmatron

Von Ardenne continued to develop the plasmatron and in 1956 he invented the Duoplasmatron, shown in Fig. 9. It is effectively the same as a plasmatron but the conical intermediate electrode is made of soft iron. The plasma chamber is positioned inside a solenoid. The conical soft iron intermediate electrode squeezes the axial magnetic field lines and concentrates them just in front of the anode. The squeezing of the magnetic field lines and the funnelling effect of the cone create a very high plasma density just in front of the extraction hole. This greatly increases the ion density and allows very high positive ion currents of up to 1.5 A to be extracted. The ions streaming through the anode hole are too dense to allow the extraction of ion beams with uniform distribution and low emittance so the plasma is allowed to expand into an expansion cup before being extracted. To prevent back-streaming electrons a suppressor electrode is used. The name duoplasmatron comes from the two (duo) different plasma densities that exist in the source.

The duoplasmatron is probably one of the most common types of positive ion source because it makes very high beam currents, is cheap and easy to maintain and works with a wide range of gases. Lifetimes are limited to a few weeks at high currents and duty factors or with heavy ions because of filament sputtering. Filaments can be easily replaced. At lower currents lifetimes can be much longer.

CERN have used a duoplasmatron on LINAC2 for over 30 years which now ultimately fills the Large Hadron Collider (LHC). It reliably produces 300 mA beams of protons in pulses up to 150 μs long at 1 Hz with lifetimes stretching to several months.

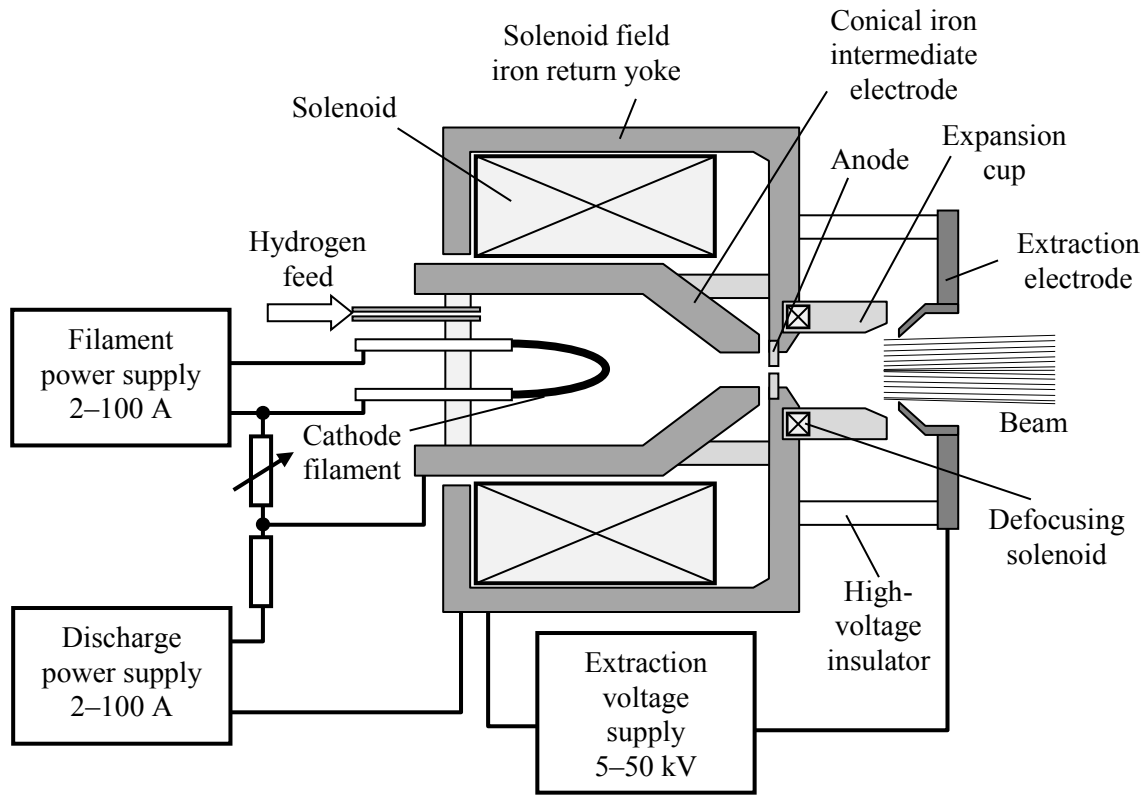

**Fig. 9:** A schematic of a duoplasmatron source

### 4.4 Microwave ion sources

#### 4.4.1 Introduction

Microwave ion sources use alternating electric fields in the gigahertz (GHz) frequency range to generate the plasma. Instead of using electrodes the microwave energy is coupled to the discharge via a waveguide. With no electrodes to erode away microwave ion sources can have lifetimes in excess of 1 year. The plasma chamber has similar dimensions to the wavelengths of the microwaves and is surrounded by DC solenoids that produce an axial magnetic field.

Microwave ion sources can be separated into two families: "on resonance" and "off resonance".

#### 4.4.2 On resonance (ECR sources)

On resonance sources are called ECR sources. The electrons are cyclotron accelerated by the combination of microwave frequency electric fields and static magnetic fields. The magnetic field makes electrons gyrate around at a frequency that matches the frequency of the microwave electric field. The solenoidal magnetic field also acts to confine the positive ions.

ECR ion sources were first developed in the late 1960s by Richard Geller and his group at CEA. ECR ion sources are very good at producing multiply charged positive ions. ECR sources work by step-wise ionization: the accelerated electrons progressively remove the outer orbital electrons of the ions by impact ionization. The comparatively slow moving positive ions are confined by the magnetic field to be ionized again by the re-accelerated electrons. High-charge-state positive ions can be produced by this technique. This is particularly useful for making high-charge-state beams of heavy elements such as uranium.

The ECR ion source is based on plasma heating at the electron cyclotron frequency ($f_{ECR}$) in a magnetic field, given by

$$\omega_{ECR} = 2\pi f_{ECR} = \frac{eB}{m} \tag{13}$$

For a 2.45 GHz frequency the electron ECR field is 875 G.

For electrons in a magnetic field in the range 0.05–1 T, this corresponds to a frequency range of 1.4 GHz to 28 GHz. The availability of commercial magnetrons and klystrons results in most sources working at 2.45, 10, 14.5, 18, 28 and 37.5 GHz.

The 2.45 GHz frequency is used because of this is also the frequency used in microwave ovens, so cheap reliable magnetron tubes are readily available. Also the waveguides are of manageable size (35 mm × 73 mm). Lower frequencies yield lower emittance beams and require lower magnetic fields.

Since 1994 CERN have used a 14.5 GHz ECR ion source to produce 100 eμA of Pb$^{27+}$ ions. Daniela Leitner and her team at Lawrence Berkeley National Laboratory have recently produced 200 eμA beams of U$^{34+}$ ions and 4.9 eμA beams of U$^{47+}$ ions with the 28 GHz superconducting VENUS ion source [3].

### 4.4.3 Off resonance (Microwave discharge sources)

Off resonance sources are called microwave discharge ion sources. They also use microwaves to produce a discharge, but the magnetic field is above the ECR field for the applied microwave frequency. Microwave discharge ion sources produce high currents of singly charged ions with low emittance. Higher plasma density is obtained by the higher magnetic fields rather than using higher gas pressure. Microwave discharge ion sources were first developed by Noriyuki Sakudo's team at Hitachi [4] and Junzo Ishikawa's team at Kyoto University [5] in the late 1970s and early 1980s. The basic design of all modern microwave discharge ion sources are based on the proton source developed by Terence Taylor and Jozef Mouris at Chalk River National Laboratory in the early 1990s [6].

### 4.4.4 Basic design

Figure 10 shows a schematic of a microwave ion source. The key aspects of the design are a small compact plasma chamber with two solenoids at the front and back. A stepped matching section is used to allow smooth transition between the waveguide and plasma impedances. An extraction system with a suppressor electrode is employed to limit the back-streaming electrons.

The main difference between ECR and microwave discharge sources is that ECR sources have an additional hexapole field surrounding the plasma chamber. This helps to further confine the ions so that high charge states can be obtained. The hexapole field can either be produced by permanent magnets or with coil windings.

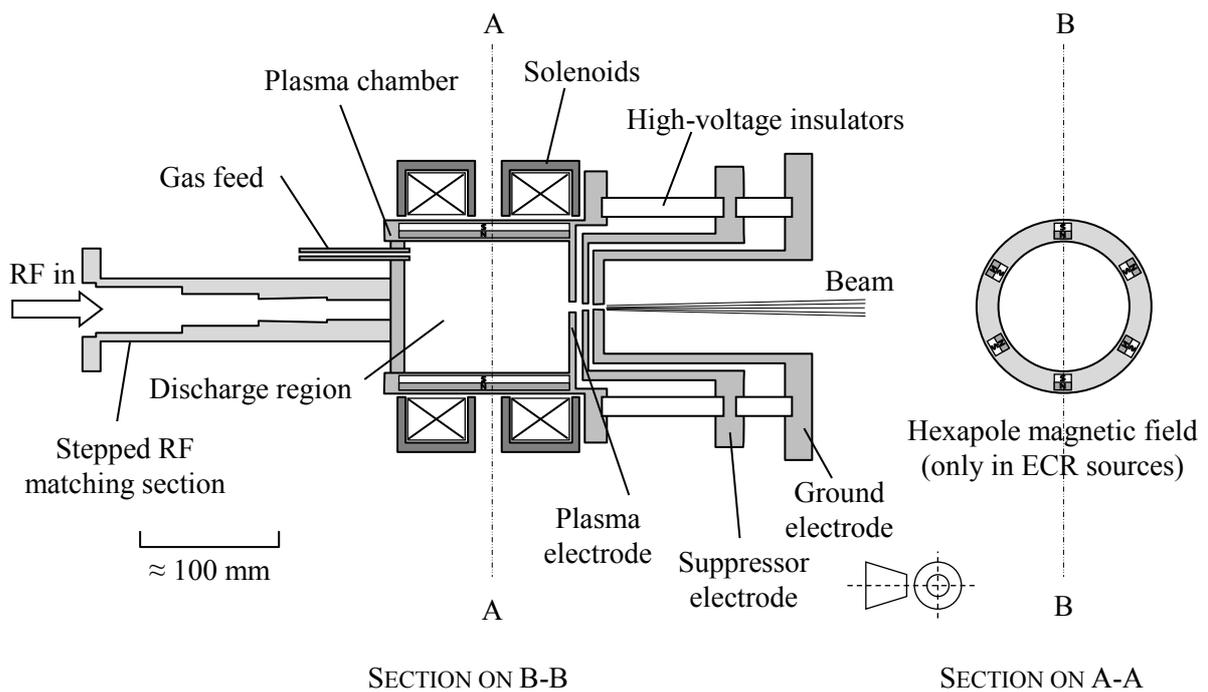

**Fig. 10:** Sectional schematic of a microwave ion source

### 4.4.5 *Further developments*

#### 4.4.5.1 *Low Energy Demonstration Accelerator*

The design was further improved by Joe Sherman and his team at LANL in the mid 1990s [7]. They optimized the extraction system and the LEBT to maximize transmission to the radio-frequency quadrupole (RFQ) of the Low Energy Demonstration Accelerator (LEDA) project. They also developed a pulsed mode operation with a rise/fall time of the order of tens of microseconds. The LEDA project demanded a decrease in beam emittance and a high reliability. The LEDA source could reliably deliver enough current to produce a 100 mA, 7 MeV DC beam of protons at the exit of the RFQ.

#### 4.4.5.2 *Source d'Ions Légers à Haute Intensité*

During the second half of the 1990s, Raphael Gobin and his team at CEA Saclay developed the Source d'Ions Légers à Haute Intensité (SILHI) source [8, 9]. They obtained greater brightness and even higher reliability. DC proton beam currents of 140 mA with 0.2 $\pi$·mm·mrad normalized emittance were achieved. Lifetimes of around 1 year were demonstrated.

### 4.5 Electron beam ion sources

#### 4.5.1 *Introduction*

EBISs use a high current density electron beam to ionize the particles. The EBIS was first developed in the late 1960s by E.D. Donets and his team at JINR, Dubna. Reinard Becker and his team at Frankfurt demonstrated that DC beams are possible but only with very low beam currents. EBISs are complex, expensive and can only produce relatively short pulse lengths of high currents. Nevertheless, they are capable of reliably producing beams of positive ions with very high charge state. Heavy elements can be completely stripped of their electrons leaving bare nuclei.

### 4.5.2    Basic operation

Figure 11 shows a schematic of an EBIS. A high-current electron gun produces a 1 keV to 20 keV electron beam that is compressed to a current density of the order of 1000 A cm$^{-2}$. The electron beam passes though through a set of drift tubes in a 1–5 T solenoidal field. The strong solenoidal field compresses the electron beam. Electrical damping components on the drift tubes help maintain the beam stability.

The material to be ionized is either pulsed into the middle of the ionization chamber or injected as a low-energy, low-charge-state beam from another ion source. The strong space charge of the negative electron beam creates a potential well that traps the injected positive ions. The amount of charge that can be trapped is limited by the size of the potential well created by the electron beam. Once trapped the ions undergo successive ionizations by the electron beam. During the trapping and ionization phase greater positive voltages are applied to the end drift tubes to longitudinally confine the ions as shown in Fig. 11. Once the required charge state has been reached the extraction phase begins by modifying the potential distribution on the drift tubes as shown in Fig. 11. The EBIS has been developed by many researchers, most recently by Jim Alessi and his team at BNL to produce a 1.7 emA, 10 µs, 5 Hz beam of Ag$^{32+}$ ions [10].

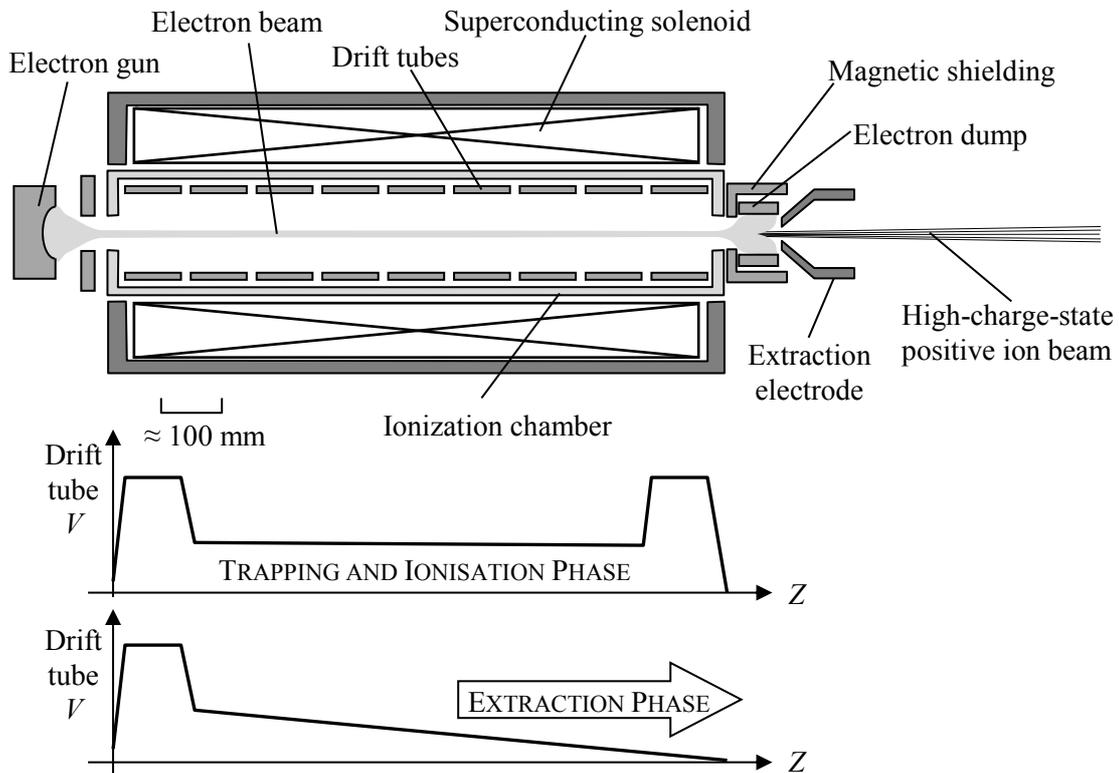

**Fig. 11:** Schematic of an EBIS

## 4.6    Laser ion sources

### 4.6.1    Introduction

Laser ion sources use a powerful laser to vaporize and ionize target material. They can produce high-current and high-charge-state beams of almost every element. The idea for laser ion sources was first proposed independently in 1969 by Peacock and Pease at UKAEA Culham and by Byckovsky and colleagues in Russia.

Laser ion sources are limited to short pulse lengths and low repletion rates. The particles are ablated from the target material so a fresh area of the target surface must be exposed for each pulse.

### 4.6.2    Basic operation

The beam from a pulsed high-power laser is focused at a target through a KCl salt window in the target chamber as shown in Fig. 12. When the laser beam hits the solid target it first vaporizes the material then ionizes it into a plasma. The electrons are accelerated by inverse bremsstrahlung to several hundred electronvolts. The electrons stepwise ionize the target atoms to higher and higher charge states. The dense plasma rapidly expands into a plasma plume which propagates along the expansion region until it reaches the extraction aperture. The target chamber sits on a high-voltage platform to allow a beam to be extracted by a grounded electrode. A suppressor electrode is also used to prevent back-streaming electrons.

The beam current produced depends on the amount of target material ionised which depends on the amount of energy delivered by the laser, this can range from 0.1 J to a few tens of Joules per pulse. The highest charge state obtained depends on the power density on the target surface. Power densities employed range between $10^9$ W/cm$^2$ and $10^{16}$ W/cm$^2$. The pulse length depends on the length of the expansion region.

The laser ion source for the TWAC at ITEP Moscow produces 7 mA, 10 µs pulses of C$^{4+}$ at 50 keV/u [11]. Recently, Masahiro Okamura at BNL has successfully matched a 35 mA, 2.1 µs pulsed beam of C$^{4+}$ from a laser ion source directly into a RFQ.

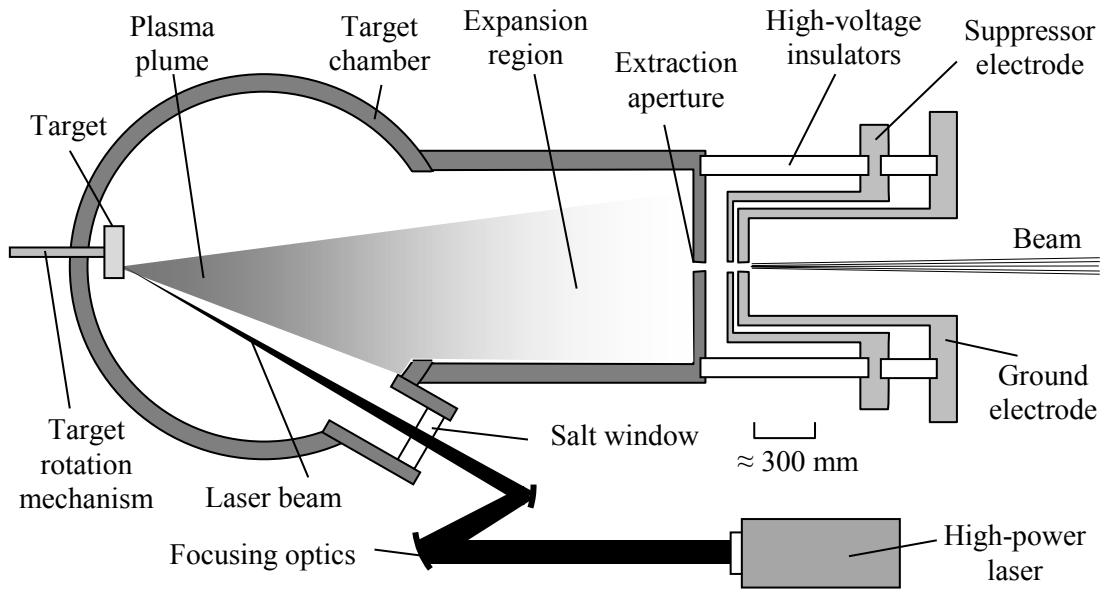

**Fig. 12:** Schematic of a laser ion source

## 4.7    Vacuum arc ion sources

### 4.7.1    Introduction

When an arc occurs in a vacuum the current-carrying particles are created by vaporizing the cathode material. Vacuum arc ion sources exploit this to produce a beam of particles made of the cathode material. Vacuum arc sources are often called metal vapour vacuum arc (MEVVA) sources. The first reliable sources were developed in the 1980s in the USA by Ian Brown, S. Humphries, Jr and

S. Picraux. MEVVA sources can produce high currents of medium-charge-state metal ions but the beam can be quite noisy. Cathode lifetimes are limited to about 1 day or less depending on the duty cycle.

### 4.7.2    Basic operation

The arc is triggered by applying a short (≈ 10 μs), high-voltage (≈ 10 kV) pulse to the trigger electrode (shown in Fig. 13). This initiates an arc between hot spots on the cathode and the anode. Microscopic irregularities on the cathode surface emit large quantities of electrons which cause localised heating. Material is vaporized from these cathode hot spots which feeds into the arc discharge. Each tiny (1–10 μm) cathode hot spot carries about 10 A and is only active for a few tens of nanoseconds before it explodes. In a typical 100–300 A arc discharge, dozens of hot spots are active at any one moment and the overall behaviour is dynamic and extremely complex.

The arc plasma expands through the expansion region which sometimes has a solenoidal field to confine the ions. The dynamic cathode hot spots make the source intrinsically noisy. Some sources use electrostatic grids [12] to stop the flow of plasma electrons so that only ions are allowed to continue to the extraction aperture. The space charge of the ions helps to smooth out the plasma density variations caused by the hot spot explosions before the ions reach the extraction aperture. The source sits on a high-voltage platform (up to 50 kV) to allow a beam to be extracted by a grounded electrode. A suppressor electrode is also used to prevent back-streaming electrons.

The MEVVA ion source for the High Current Injector at GSI in Germany can produce 15 mA of $U^{4+}$ ions [13].

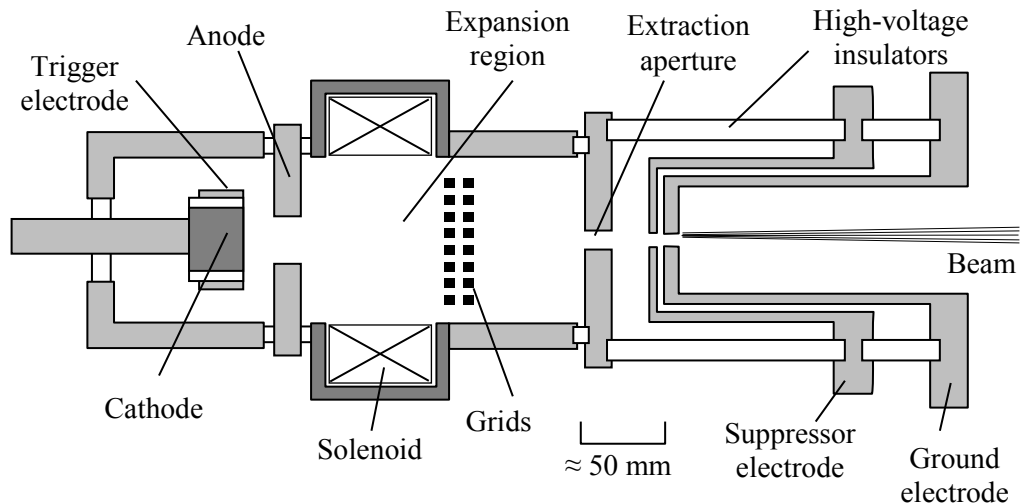

**Fig. 13:** Schematic of a vacuum arc ion source

## 5    Negative ion sources

### 5.1    Introduction

#### 5.1.1    The negative ion

Negative ion sources produce beams of atoms with an additional electron. The binding energy of the additional electron to an atom is termed the electron affinity. Some elements have a negative electron affinity (such as beryllium, nitrogen or the noble elements) which means they cannot form stable negative ions. H⁻ is the most commonly produced negative ion. Hydrogen has an electron affinity of 0.7542 eV. Considering that the electron binding energy of neutral hydrogen is 13.6 eV, the extra electron on an H⁻ ion is very loosely held on. All of the ion sources in this section have been used to produce D⁻ ions as well as other heavy negative ions, such as O⁻, B⁻, C⁻, etc.

### 5.1.2    Uses

Negative ion sources were first developed to allow electrostatic accelerators to effectively double their output beam energy. In a tandem generator H⁻ ions are first accelerated from ground to terminal volts, they are then stripped of their two electrons when they pass through a thin foil. The resulting protons are then accelerated from terminal volts back to ground, at which point they have an energy of twice the terminal volts.

Cyclotrons use negative ions and stripping foils to extract the beam from the cyclotron. The stripping foil is positioned near the perimeter of the cyclotron poles. As the negative ion beam is accelerated it circulates on larger and larger radii until it passes through the stripping foil, which converts the beam from being negative to positive. The Lorenz force on the beam is reversed and instead of the force pointing into the centre of the cyclotron it points outwards and the beam is cleanly extracted.

In high-power proton accelerators H⁻ ions are used to allow charge accumulation via multiturn injection. An H⁻ beam from a linear accelerator is fed through a stripping foil into a circular ring (a storage, accumulator or synchrotron ring) leaving protons circulating in the ring. The H⁻ beam from the linear accelerator continues to enter the ring whilst the circulating beam repeatedly passes through the stripping foil unaffected. The incoming beam curves in one way through a dipole as an H⁻ beam then curves out of the dipole in the opposite direction as a proton beam on top of the circulating beam. This allows accelerator designers to beat Liouville's theorem and build up charges in phase space. Without this negative ion stripping trick only one turn could be accumulated in the ring.

## 5.2    Physics of negative ion production

### 5.2.1    Mechanisms and challenges

The physical processes involved with the production of negative ions are still not fully understood, but they can be generally described as: charge exchange, surface and volume production processes. In different types of source one production process may dominate, however all three processes might contribute to the overall extracted negative ion current.

### 5.2.2    Charge exchange

The first H⁻ ion sources were charge exchange devices. There are two ways of doing this: with foils or gases. With foils a proton beam, with an energy of about 10 keV, is passed though a negatively biased foil and by electron capture an H⁻ beam is produced. For gases the proton beam is passed though a region filled with a gas. The H⁻ beam is produced by sequential electron capture; first protons are converted to neutral H⁰, then to H⁻. The gas acts as an electron donor. Only about 2% of the protons are converted into H⁻ ions. Until the 1960s this was the main technique used to make H⁻ beams. Beams of up to 200 μA were produced using this method. In 1967 Bailey Donnally [14] discovered that the yield of He⁻ ions can be increased by using caesium vapour as an electron donor. This lead to the development of a series of negative ion sources using alkali vapour.

Resonant charge exchange between fast H⁻ ions and slow neutral hydrogen atoms (H⁰) is essential to the operation of a Penning source (see Section 5.3.3):

$$H^- (\approx 60 \text{ eV}) + H^0 (< 1 \text{ eV}) \rightarrow H^- (< 1 \text{ eV}) + H^0 (\approx 60 \text{ eV})$$

### 5.2.3    Early Negative Ion Sources

For several decades numerous researchers [15, 16] had been experimenting with sources originally designed to produce positive ions, but by reversing the polarity of the extraction they were able to extract negative ions. Nevertheless, the co-extracted electron current was always at least an order of magnitude higher than the negative ion current.

In the early 1960s George Lawrence and his team at Los Alamos [17] were using a duoplasmatron to produce H$^-$ ions when they first noticed that substantially higher beam currents and lower electron currents could be extracted when the extraction was actually off-centred from the intermediate electrode (Fig. 14). They concluded that the extracted H$^-$ ions must be produced near the edge of the plasma. (This was also discovered independently by a team at the UK Atomic Weapons Establishment [18].)

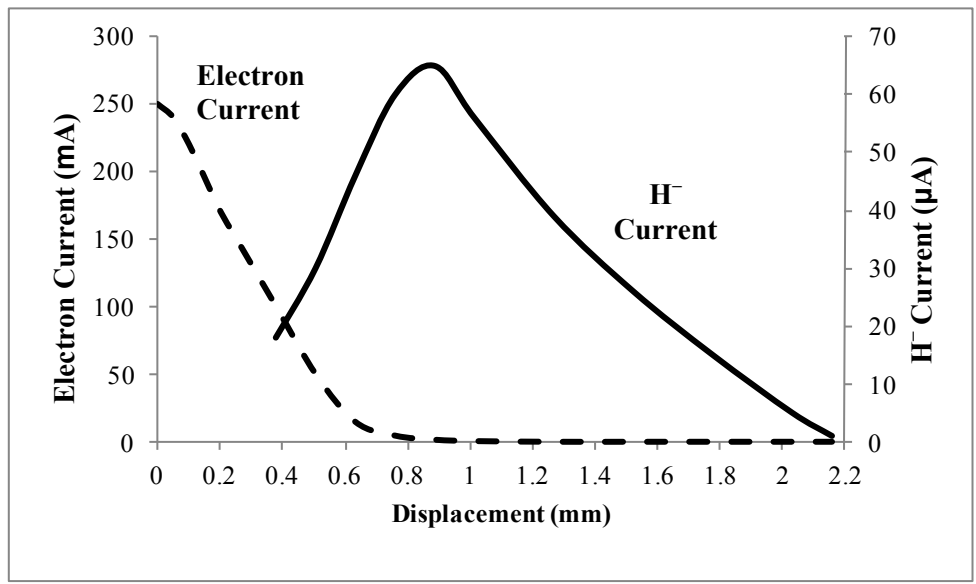

**Fig. 14:** H$^-$ and electron currents as a function of extraction offset in a duoplasmatron measured at Los Alamos

During the 1960s various sources originally designed to produce positive ions were adapted and modified to produce H$^-$ ions and beam currents up to a few milliamps were produced.

### 5.2.4    Caesium and surface production

In the early 1960s Victor Krohn, Jr and his team [19] at Space Technology Laboratories, Inc. California were experimenting with surface sputter ion sources. Surface sputter ion sources are mainly used to produce beams of heaver ions (such as metals) for coating and etching applications. Krohn noticed that when Cs$^+$ ions were used to sputter a metal target the yield of sputtered negative ions increased by an order of magnitude.

In the early 1970s Gennadii Dimov, Yuri Belchenko and Vadim Dudnikov at the Budker Institute of Nuclear Physics started experimenting with caesium in ion sources. Using a magnetron ion source (see Section 5.3.2), Vadim Dudnikov added Cs vapour to the discharge for the first time. A dramatic increase in H$^-$ current was observed along with a decrease in co-extracted electrons. The Dimov team went on to extract a colossal 880 mA pulsed H$^-$ beam from an experimental magnetron ion source [21]. This success led them to develop a Penning type ion source (see Section 5.3.3) that could produce 150 mA of H$^-$ beam current with only 250 mA of extracted electrons. The H$^-$ currents produced were orders of magnitude higher than anything seen previously. When these revolutionary results were published interest in caesiated ion sources took off. Researchers all over the world started using caesium in their ion sources and a large number of new ion source designs were developed.

A very different type of H⁻ source that relies on surface production of H⁻ ions is the surface converter source (see Section 5.5). Developed in the 1980s by Ehlers and Leung at the Lawrence Berkeley Laboratory, it also relies on a caesiated surface. The caesiated surface sits in the middle of the plasma and is curved with a radius centred on the extraction region. H⁻ ions produced on this surface are "focused" towards the extraction hole because they are repelled by the negative potential on the converter surface, this is why this type of source is also called "self-extracting".

In addition to aiding H⁻ surface production, caesium also helps to stabilize the plasma by readily ionizing to produce additional electrons for the discharge. This reduces the amount of noise in the discharge and extracted beam current.

### 5.2.5    Surface physics processes

The types of particles arriving at a surface could be protons, ionized hydrogen molecules, ionized caesium, energetic neutral atoms or molecules. When a particle interacts with a surface many complex and competing processes can occur:
- reflection;
- adsorption;
- sputtering;
- desorption;
- recombination;
- dissociation;
- ionization;
- secondary electron emission;
- photo emission;
- excitation.

Particles ejected from the surface could be of a different charge state from the incoming particle, be excited, be in a molecule, or some combination of all three. The surface may also be altered. Complex interactions can take place at the surfaces.

The most important factor affecting H⁻ production at a surface is the work function, φ. To make H⁻ ions the surface must provide electrons, so a low work function surface is essential. The work function of a surface obviously depends on what it is made of. If different atoms of a different element are adsorbed on that surface (such as caesium) then the work function can be altered. The "thickness" of the adsorbed layer will also have an effect on the surface's work function.

The thickness of the adsorbed layer is usually defined in terms of the number of "monolayers" of the adsorbed atoms:

$$\text{Thickness (number of monolayers)} = \frac{\text{Number of adsorbate atoms per unit area}}{\text{Number of adsorbate atoms for a monolayer per unit area}} \qquad (14)$$

When talking about negative ion production, the surface is usually the cathode and is typically made of a high melting point metal such as tungsten φ = 4.55 eV or molybdenum φ = 4.6 eV. Caesium has the lowest work function of all elements: φ = 2.14 eV. The work function of a caesium-coated molybdenum surface is actually lower than that of bulk caesium. As caesium covers the molybdenum surface the work function decreases to 1.5 eV at 0.6 of a monolayer and then rises to about 2 eV for one monolayer or greater of caesium as shown in Fig. 15. This minimum at 0.6 monolayers is caused by atomic interactions increasing the Fermi level at the surface, thus decreasing the amount of energy required to liberate the electrons.

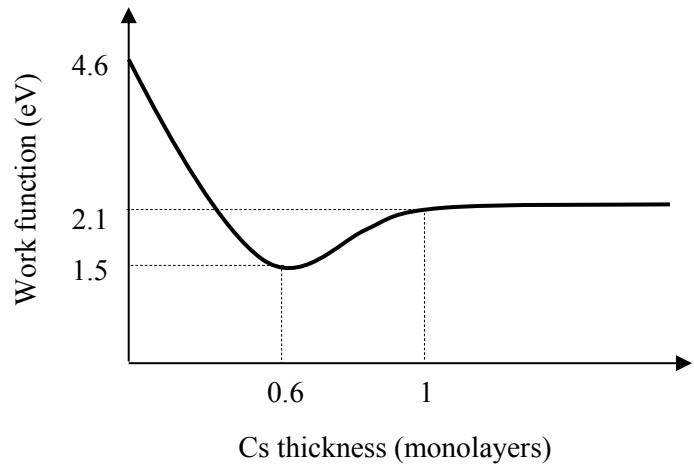

**Fig. 15:** Surface work function versus caesium thickness on a molybdenum surface

The thickness of the caesium coating on electrode surfaces will rarely ever come close to one full monolayer because thermal emission and plasma sputtering removes excess caesium. This is because the Cs–Cs bond is actually weaker than the Cs–Mo or Cs–Ta bonds, so Cs atoms adsorbed on Cs atoms are rapidly sputtered away by the plasma or thermally emitted from hot surfaces. (It is possible to build up multiple layers but only on cold surfaces that are shielded from the plasma.)

### *5.2.6    Maintaining caesium coverage*

To minimize the work function and maximize the H⁻ production an optimum layer of caesium must be maintained on the surface. The surface is a dynamic place, caesium atoms are constantly being desorbed by plasma bombardment. To maintain optimum caesium coverage a constant flux of caesium is required. This is often provided by an oven containing pure elemental caesium but it can also be provided using caesium chromate cartridges that release Cs when heated.

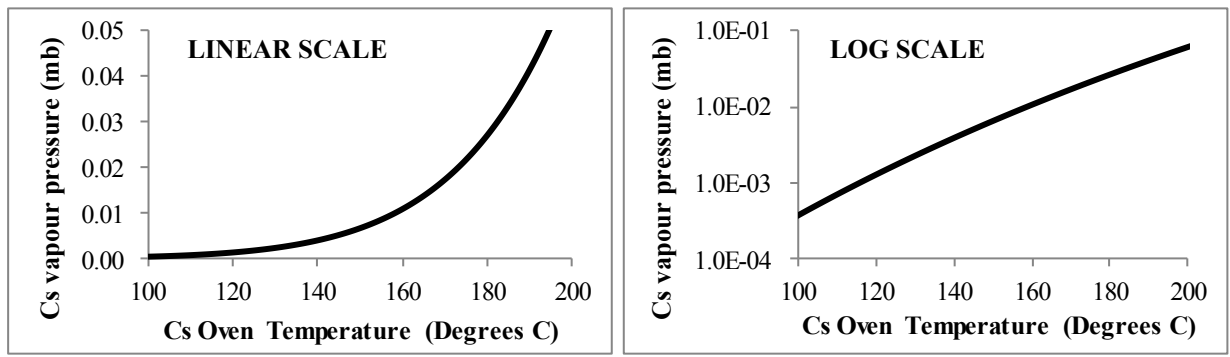

**Fig. 16:** Caesium vapour pressure versus caesium oven temperature

The flux of caesium into the plasma can be precisely controlled by setting the temperature of the caesium oven. Figure 16 shows how the vapour pressure of caesium varies with temperature [20]. All sources that use elemental caesium in an oven operating between 100 °C and 190 °C. This covers a large range of vapour pressures. Most sources operate in the 130 °C to 180 °C range but some sources such as the RF driven volume multicusp source only require very small Cs fluxes and operate in the 105 °C to 110 °C range.

### 5.2.7　Volume production

Also in the 1970s in parallel to the discovery of caesium-enhanced H⁻ surface production, Marthe Bacal and her team at Ecole Polytechnique developed a completely new type of source that relied on H⁻ production in the plasma volume itself. Initially people were sceptical because H⁻ ions are so fragile: only 0.7542 eV is required to detach the extra electron. The plasma in the discharge was thought to be too energetic for any H⁻ ions produced in the volume to survive long enough to make it to the extraction region. The breakthrough that made volume production possible was separation by a magnetic filter field of the plasma production region from the extraction region. In the 1980s, Leung, Ehlers and Bacal used a filament-driven multicusp ion source with a magnetic dipole filter field positioned near the extraction region (see Section 5.6). The filter field blocked high-energy electrons from entering the extraction region, whereas ions and cold electrons could diffuse across the filter field. This effectively separated the discharge into two regions: a high-temperature driver plasma on the filament side of the filter field, and a low-temperature H⁻ production plasma on the extraction region side. Magnetically filtered multicusp ion sources are sometimes called "tandem" sources because of these two regions of different plasma temperatures (not to be confused with tandem accelerators).

The volume production process relies on the dissociative attachment of low energy electrons to rovibrationally excited $H_2$ molecules:

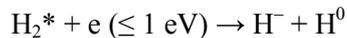

$$H_2{}^* + e\ (\leq 1\ \mathrm{eV}) \rightarrow H^- + H^0$$

If the $H_2$ molecule is vibrationally cold the dissociative attachment cross section is extremely low ($10^{-21}$ cm$^2$). When the $H_2$ molecule is rovibrationally excited, however, the cross section increases by five orders of magnitude. Thus, low-energy electrons can be very effective in generating H⁻ ions by dissociative attachment to highly vibrationally excited molecules. The rovibrationally excited molecules are produced not only in the plasma but also on the walls of the chamber and electrode surfaces.

### 5.2.8　H⁻ destruction

In both volume and surface sources there are many processes that can destroy H⁻ ions. The most common are:

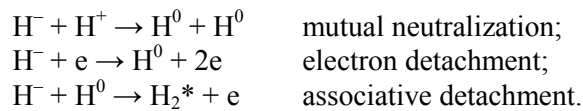

$$H^- + H^+ \rightarrow H^0 + H^0 \quad \text{mutual neutralization;}$$
$$H^- + e \rightarrow H^0 + 2e \quad \text{electron detachment;}$$
$$H^- + H^0 \rightarrow H_2{}^* + e \quad \text{associative detachment.}$$

Another important factor is the cross section of the H⁻ ion in comparison to the H⁰ atom. The H⁻ ion cross section is 30 times larger than the neutral H⁰ atom for collisions with electrons and 100 times larger for collisions with H⁺ ions. As well as being fragile and easily destroyed it is much more likely to be hit.

The aim of the source designer is to minimize the H⁻ destruction processes by controlling the geometry, temperature, pressure and fields in the source. The following sections describe the design of some of the most successful H⁻ ion sources.

## 5.3　Surface plasma cold cathode ion sources

### 5.3.1　Introduction

The term cold cathode refers to the fact that the cathode is not independently heated, however the name can be misleading because the cathode can still operate at elevated temperature due to heating by the discharge itself. They are called surface plasma sources because H⁻ ions are produced on the surface of the cathode (see Section 5.2.5). Both of the sources discussed in this section were used for many years as positive ion sources before they were employed to make H⁻ ions. The discharge is in direct contact with the anode and cathode, so sputtering processes will eventually erode the electrode surfaces. This puts a fundamental limit on their lifetime.

### 5.3.2 Magnetron

The magnetron (also called a planotron) was the first ion source where the H⁻ current was first significantly increased by adding caesium vapour. This work was done by Gennadii Dimov, Yuri Belchenko and Vadim Dudnikov at the Budker Institute of Nuclear Physics in the early 1970s [21]. Chuck Schmidt developed the Fermilab version of Magnetron ion source in the late 1970s, a design which was adopted and further developed by Jens Peters at DESY, and Jim Alessi at Brookhaven.

The magnetron source has a racetrack-shaped discharge bounded on the inside by the cathode and the outside by the anode as shown in Fig. 17. The anode and cathode are only about 1 mm apart so the discharge is in the shape of a ribbon wrapped around the cathode. A magnetic field of between 0.1 T and 0.2 T is applied perpendicular to the plane of the racetrack, this causes the plasma to drift around the racetrack. On one of the long sides of the racetrack discharge the anode has a hole through which the beam is extracted. Pulsed hydrogen is fed into discharge on the opposite side to the extraction hole. Caesium vapour is introduced via an inlet on one side. H⁻ ions are produced on the cathode surface and are accelerated away by the cathode sheath potential. A concave region on the cathode surface opposite the extraction hole can give an initial focus to the extracted H⁻ beam. Some of the cathode sheath accelerated H⁻ ions undergo resonant charge exchange with slow thermal $H^0$ on the way to extraction, resulting in a beam energy distribution with two peaks.

The magnetron can give high H⁻ currents up to 80 mA and can have very long lifetimes of over 6 months, but it will only operate at very low duty factors of up to 0.5 %. This because it is not possible to maintain optimum caesium coverage on the cathode surfaces during the time the discharge is on.

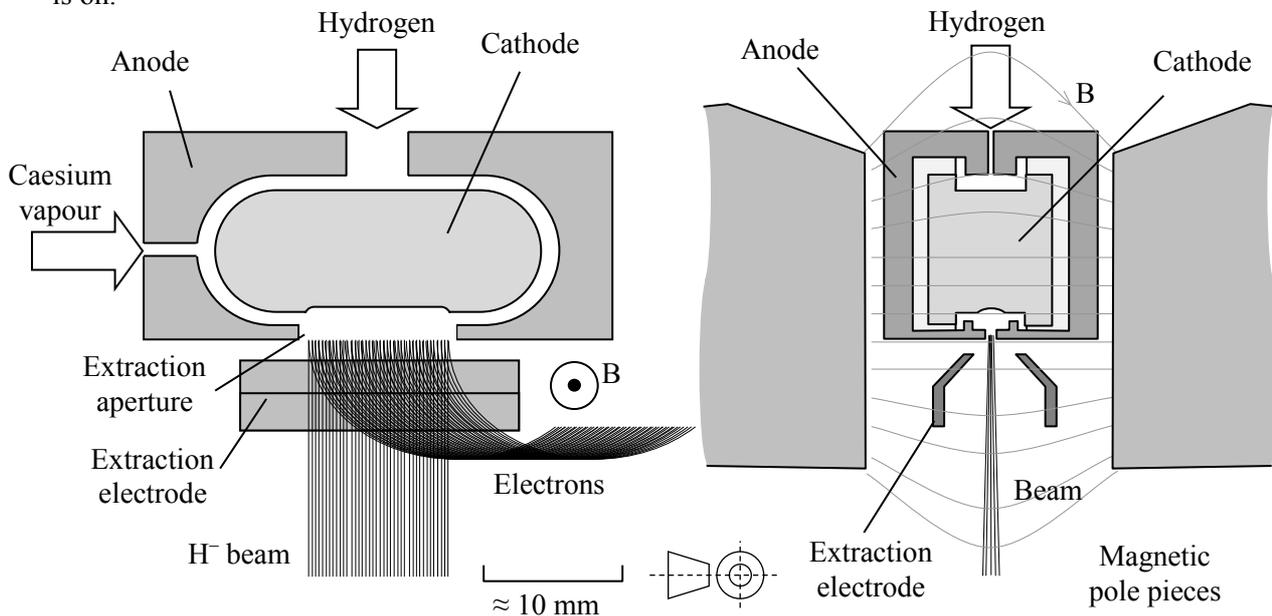

**Fig. 17:** Sectional schematic of a magnetron with slit extraction

### 5.3.3 Penning

The H⁻ Penning ion source was invented by Vadim Dudnikov, and developed with Gennadii Dimov and Yuri Belchenko at the Budker Institute of Nuclear Physics in the early 1970s at the same time as the magnetron source. Vadim Dudnikov reported high H⁻ currents up to 150 mA and duty cycles all of the way up to DC [22]. Vernon Smith, Paul Allison and Joe Sherman at Los Alamos developed a scaled up Penning source [23] that gave similarly high currents with low emittances.

A Penning source (Fig. 18) has a small (10 mm × 5 mm × 5 mm) rectangular discharge region with a transverse magnetic field. The long sides of discharge are bounded by two cathodes, with the other four walls at anode potential, creating a 'quadrupole like' electric field arrangement. The magnetic field is orientated orthogonally to the cathode surfaces so that electrons emitted from the cathode are confined by the magnetic field lines and reflex back and forth between the parallel cathode surfaces. The primary anode is hollow and has holes through which hydrogen and caesium vapour are fed into the discharge. The extraction aperture plate is also at anode potential. The beam is extracted from the plasma through a slit in the extraction aperture plate by a high-voltage applied to an extraction electrode. The electrodes are made of molybdenum.

Like with the magnetron, $H^-$ ions are produced on the cathode surfaces and accelerated by the plasma sheath potential that exists next to the cathode. The plasma sheath potential is about 60 V. Unlike the magnetron, however, the cathode surface is not directly opposite the extraction aperture. The addition of ribs on the inside of the extraction aperture plate mean there is no direct line of sight between the cathode surface and the extraction aperture. Thus, it is impossible for the fast cathode produced $H^-$ to be extracted directly. Only $H^-$ ions that have undergone resonant charge exchange with slow $H^-$ ions (see Section 5.2.2) are extracted resulting in a beam with a lower energy spread than from the magnetron.

The Penning is the brightest $H^-$ ion source with current densities at extraction above 1 A cm$^{-2}$ possible. The lifetime of the Penning source is limited to a few weeks because of cathode sputtering by caesium ions. This type of source will not operate without caesium vapour and requires the electrode surfaces to be between 400 °C and 600 °C.

This source is currently under development at the Rutherford Appleton Laboratory by the present author and his team; 60 mA, 1 ms, 50 Hz pulses can be routinely produced [24].

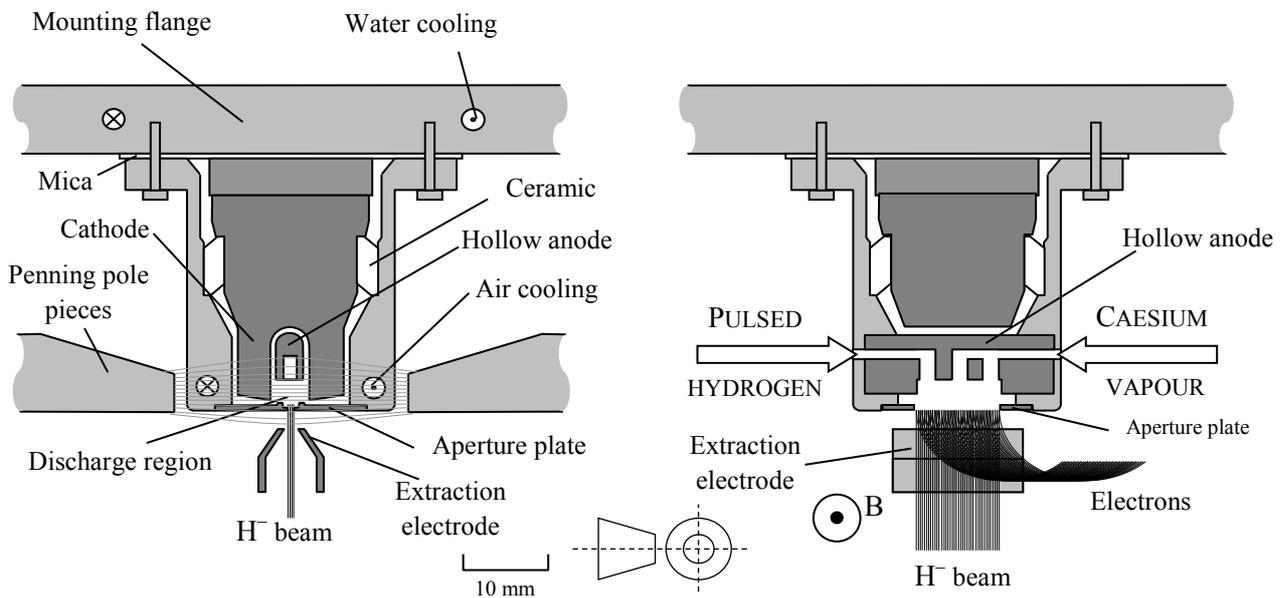

**Fig. 18:** Sectional schematic of a Penning $H^-$ ion source

## 5.4    Multicusp ion sources

### 5.4.1    Introduction

Multicusp ion sources have permanent magnets positioned around the perimeter of the plasma chamber with alternating north and south poles. This alternating arrangement creates magnetic cusps around the chamber walls which serve to confine the plasma and keep it away from the walls. The containment of the plasma by the multipole field arrangement is why multicusp sources are also referred to as "bucket" sources. Originally developed in the 1970s for fusion research these sources truly were giant buckets of plasma with dimensions of the order of 0.5 m to 1 m. They were designed to produce several amps of current extracted through multiple apertures in an extraction grid.

For high-current multicusp sources the means of plasma production is either in a hot cathode discharge or in an inductively coupled discharge with a RF solenoid antenna. Researchers have investigated both microwave and ECR discharge plasma production in multicusp sources, but only low negative ion beam currents in the microamp range have been produced, so they are not used for high-power accelerator applications.

### 5.4.2    Hot-cathode-driven plasma production

Most multicusp sources use a hot tungsten filament cathode bent into the shape of a hairpin. Using a filament heater power supply, several hundred amps of DC are passed through the filament to heat it up so that it thermionically emits electrons. The main plasma discharge is then created with a second power supply, between 10 A and 500 A DC flows from the filament cathode to an anode. The anode can either be the metallic walls of the plasma chamber or a specific electrode in the plasma chamber.

The temperature of the filament should not be too high otherwise the electron emission will become space charge limited. This causes the main discharge voltage to increase and become very noisy. The shape and position of the hot cathode filament is important. Coil filaments can be used but the hairpin shape is more common. A notable exception is the JPARC source that uses a coil filament made from $LaB_6$ as a cathode [25]. At about 1500 °C $LaB_6$ produces large amounts of electrons.

Wide hairpins (U-shaped) have been shown to produce more stable plasmas over a large range of operating conditions. The position of the filament relative to the multicusp field is important because the high current required to heat the filament loop creates a magnetic field itself. The filament should therefore be positioned so that the filament produced magnetic field is in the same direction as the local cusp field.

### 5.5    Filament cathode multicusp surface converter source

The multicusp surface converter source is a multicusp filament-driven discharge with the H⁻ ions produced on a molybdenum converter surface opposite the extraction hole. It was first developed by Ehlers and Leung and the team at Lawrence Berkeley University in the late 1970s and early 1980s. This type of source has also been used to generate other negative ions.

A discharge of several hundred amps is created between the filament cathode and the anode walls. The multicusp field as shown in Fig. 19 confines the plasma. Caesium vapour is fed into the discharge chamber and it coats the converter surface. This type of source is sometimes called a self-extracting source because: (1) the converter is negatively biased so H⁻ ions produced on its caesium-coated surface are repelled toward the extraction hole; and (2) the radius of curvature of the converter surface is centred on the extraction hole to focus the H⁻ ions towards the extraction aperture. The multicusp magnets on either side of the extraction region act as a filter field to allow some volume production of H⁻ ions to supplement the ions produced by the surface converter. They also provide the magnetic field to dump the co-extracted electrons.

Ehlers and Leung developed this source for neutral beam injectors for fusion research to produce a 1 A DC H⁻ beam by adding more cathode filaments (up to six) and increasing the discharge current to 1000 A (see Ref. [26]).

Filament-driven surface converter sources tend to consume large amounts of Cs to negate the effect of the sputtered filament atoms adsorbing on the Cs layer on the surface converter electrode.

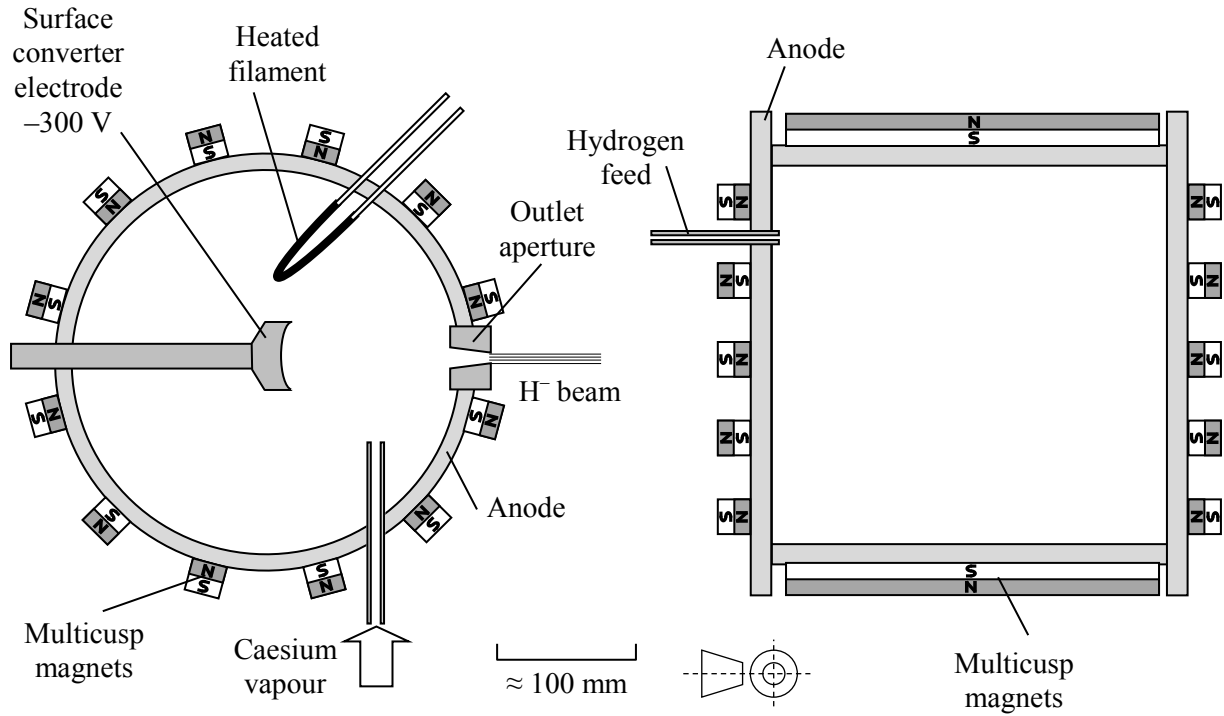

**Fig. 19:** Sectional schematic of a filament cathode multicusp surface converter source.

Rod Keller and Gary Rouleau at Los Alamos National Laboratory use this type of source operations on the LANSCE machine routinely producing a 16 mA, 60 Hz H⁻ beam with a lifetime of 35 days [27]. Like all filament driven discharge sources it suffers from lifetime limitations due to filament erosion. This source takes about 10 hours to start up and stabilize its output. It takes this long to develop the equilibrium coverage of caesium on the surface converter. This can impose operational restrictions on the rest of the machine.

## 5.6   Filament cathode multicusp volume source

The filament cathode multicusp volume source with filter field was first developed by Ehlers, Leung and Marthe Bacal and the teams at LBNL and Ecole Polytechnique in the 1980s [28]. A schematic of the source is shown in Fig. 20. The filter field creates a low-electron-temperature plasma region that is conducive to H⁻ production in the volume just in front of the extraction region (see Section 5.2.7 for an explanation of the physics).

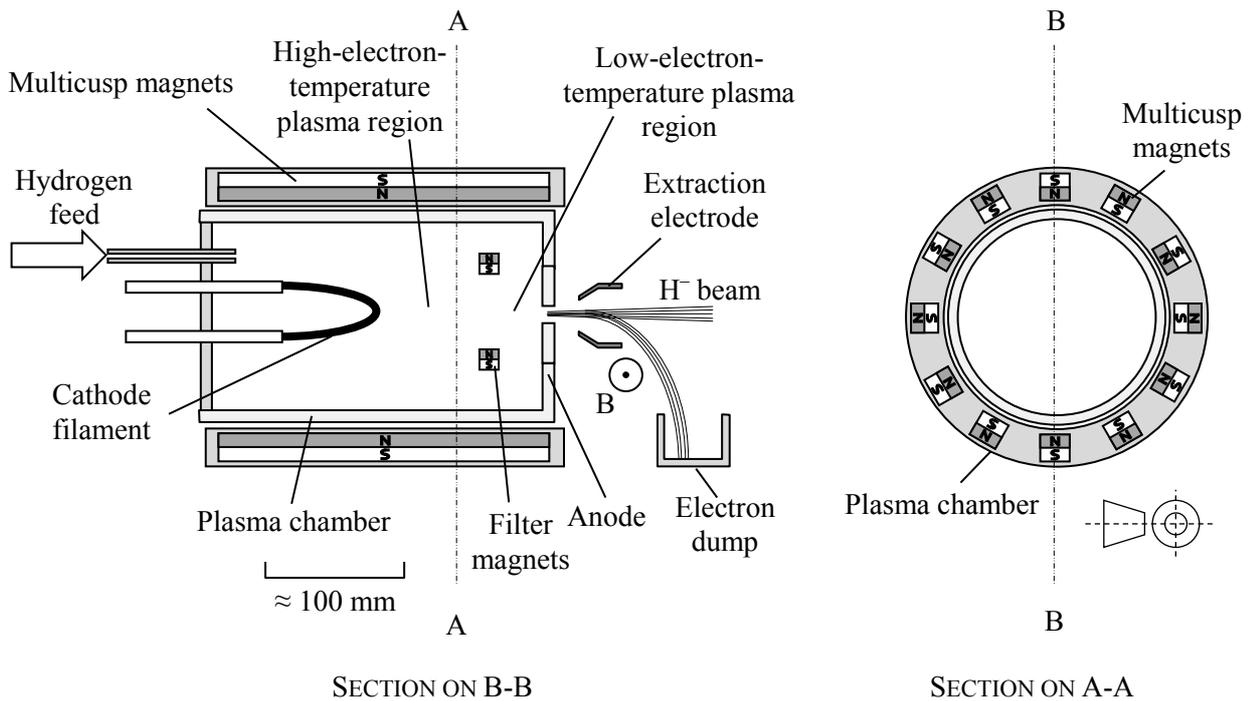

SECTION ON B-B                    SECTION ON A-A

**Fig. 20:** Schematic of a filament cathode multicusp volume source

This type of source has been successfully developed all over the world and a few companies now sell them commercially. They are reliable and although they only have a lifetime of a few weeks they are very low maintenance, only requiring a very simple filament change. They are used mainly on cyclotrons.

Andrew Holmes and his team at Culham found the maximum current that can be extracted from this type of source is about 40 mA DC; it is limited by the maximum H⁻ density obtainable at the extraction region. The H⁻ current does not increase above a discharge current of about 200 A because H⁻ destruction processes start to dominate [29].

If caesium is added to this type of source the H⁻ current can be doubled to 80 mA. The caesium does not increase the volume production of H⁻ instead it actually facilitates surface production making this source a combined volume and surface source.

Frankfurt University created a source with three cathode filaments and using caesium they produced pulsed H⁻ currents of 120 mA [30], however the emittance and persistence of this beam was never measured and it is unlikely that anywhere near that current could actually be transported to the next stage of an operational accelerator. The lifetime of the Frankfurt University source was also never fully evaluated.

## 5.7   Internal RF antenna multicusp volume source

Instead of producing a discharge between a hot cathode filament and an anode the discharge can be generated by inductively coupled RF heating. An alternating magnetic field is produced by solenoid antenna fed with a RF power supply at a frequency of between 1 MHz and 10 MHz. Figure 21 shows a schematic of the source which also includes a filer field to screen the H⁻ volume production region from fast electrons.

In the early 1990s Ka-Ngo Leung and his team at Lawrence Berkeley Laboratory first developed this type of source with a two and a half turn antenna inside the plasma chamber. The antenna was made of 4.7 mm diameter copper tubing and coated with porcelain. It was powered with a

2 MHz, 50 kW RF power supply. The antenna was water-cooled. They obtained a H⁻ current of about 40 mA which could be increased to about 90 mA by adding caesium. The lifetime is limited to a few weeks due to antenna erosion.

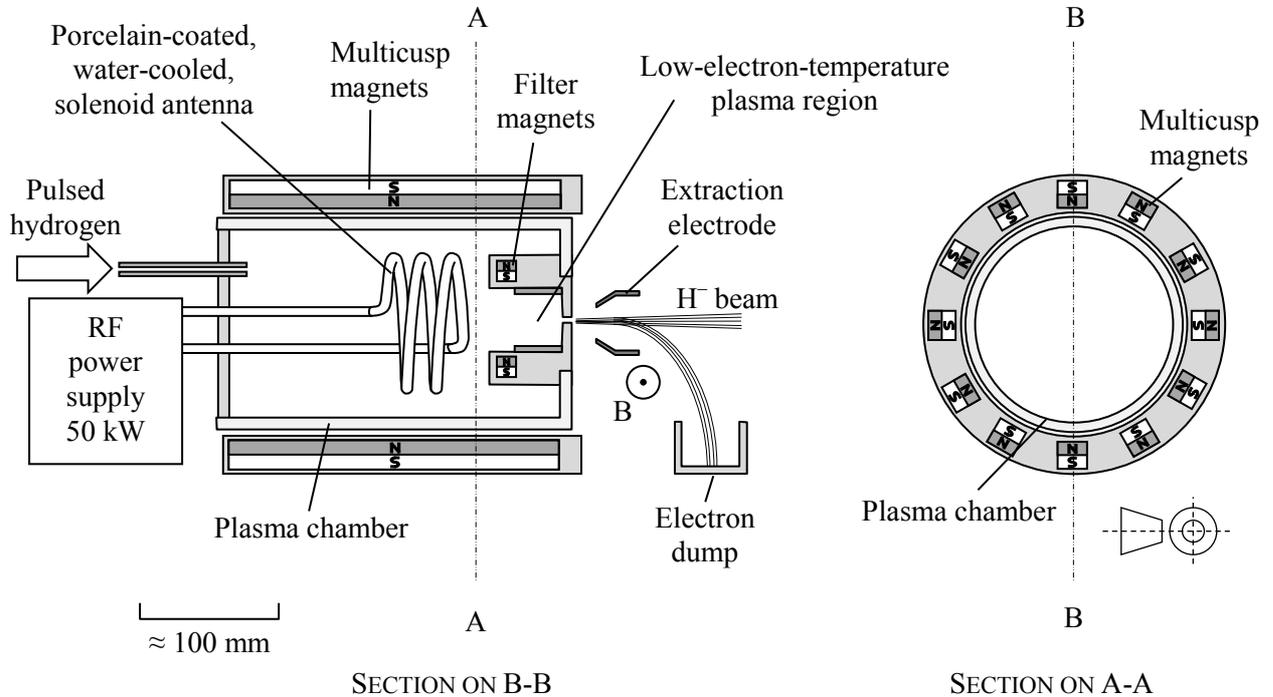

**Fig. 21:** Schematic of an internal RF antenna multicusp volume source.

In the 2000s Martin Stockli and his team at Oak Ridge National Laboratory developed this source for SNS operations. This source now routinely injects 50 mA, 1 ms, H⁻ pulses at 60 Hz into the RFQ, of which 38 mA are accelerated by the LINAC [31]. Source lifetimes are of the order of 4–5 weeks. Failures are eventually caused by hot spots occurring on the antenna which melts the 0.6 mm porcelain coating.

During start up, approximately 3 mg of caesium is introduced into the discharge by heating caesium chromate cartridges. No more caesium is added for the lifetime of the source [32].

## 5.8 External RF antenna multicusp volume source

The problem of antenna failure in multicusp sources can be avoided by putting the antenna outside the plasma chamber. Jens Peters and his team at DESY [33] were the first to successfully try this for negative ions in the late 1990s. They fed 50 kW RF into a three-turn solenoid antenna outside an $Al_2O_3$ ceramic chamber and obtained a 40 mA H⁻ beam with a duty factor of 0.05% and a pulse length of 100 μs without caesium. The source ran for a record breaking 300 days. They also experimented with different number of turns on the antenna and different frequencies ranging from 1.65 MHz to 9 MHz. The optimum appeared to be about five turns and 2 MHz.

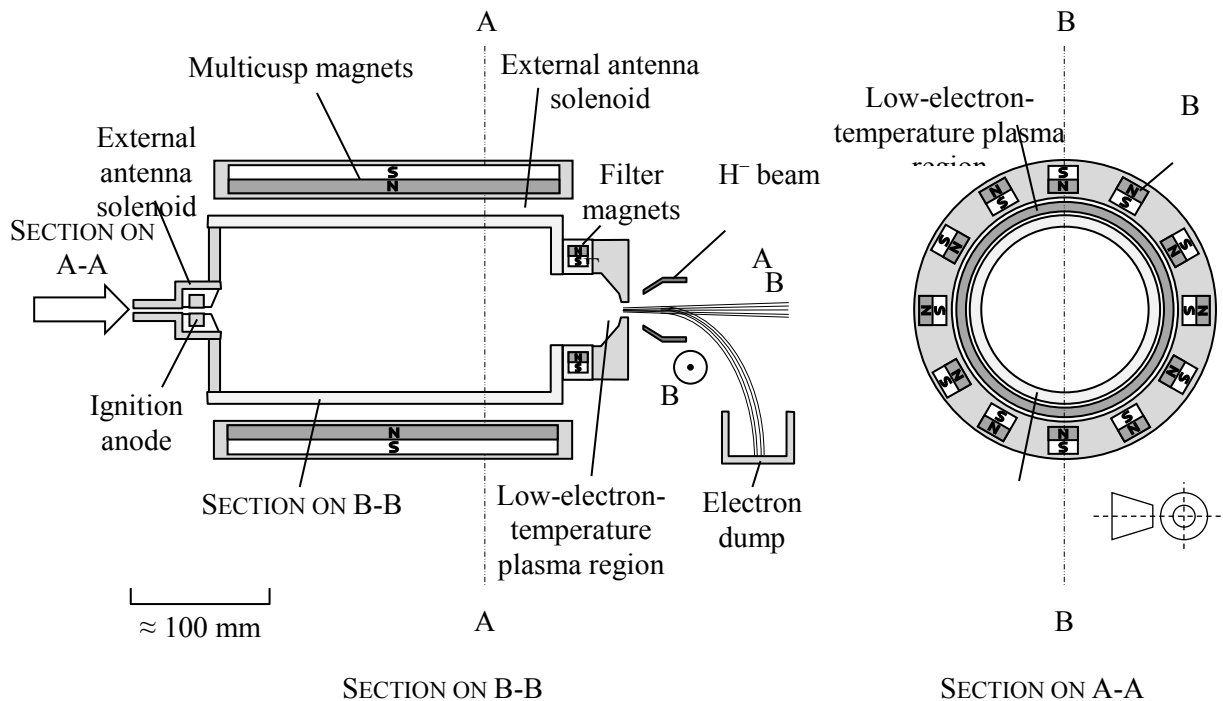

**Fig. 22:** Sectional schematic of an external antenna RF multicusp volume ion source

This epic lifetime inspired Oak Ridge National Laboratory to start developing an external antenna source for the SNS in the mid 2000s. The DESY source ran at very low repetition rate and duty cycle. SNS requires 1 ms beam pulses at 60 Hz: two orders of magnitude greater. When they tried to scale up the duty cycle they found that they could not extract high enough beam currents. Rob Welton and the SNS ion source group are currently developing this source. Experiential caesiated sources on a test rig have demonstrated unanalysed currents up to 100 mA (see Ref. [34]), but lacked persistence at the required duty factor. External helicoil and saddle antennas have been tested in an attempt to increase the plasma density near the extraction region.

In the late 2000s Jacques Lettry and his team at CERN also started developing an external solenoid antenna multicusp plasma generator. As of 2011 a stable 2 MHz, 1.2 ms, 50 Hz plasma has been produced, but extraction has yet to be demonstrated [35].

Figure 22 shows a five-turn external solenoid antenna multicusp volume source. The pulsed hydrogen is pre-ionized by a spark gap on injection into the plasma chamber. A filter field is positioned in front of the extraction aperture to provide a volume production region.

# 6    Running and developing sources

## 6.1    Which source?

The type of source an accelerator uses obviously depends on what types of ions are required and what beam current is needed. The reason why one type of source in a family is used and not another is usually historic. It depends on when the accelerator was built, what facility was there previously and what expertise is available. Cost can sometimes be an issue as well.

CERN use a duoplasmatron because they need a 300 mA proton current. Fermilab use a magnetron because it was the best source that met their needs when it was developed in the late 1970s.

Similarly RAL uses a Penning source because it was the best source that met their needs in the early 1980s.

If you were building a new machine today which source would you choose? For proton sources up to 100 mA the microwave discharge ion source is the obvious choice because it offers exceptional lifetime and reliability. For proton sources up to 500 mA, the only option is the duoplasmatron. For heavy or multiply charged ions the ECR ion source is the best option. For high-charge-state ions, the EBIS is the best option.

The question is much more interesting when considering negative ion sources. Intense development work continues at all of the major labs operating $H^-$ ion sources.

External RF antenna multicusp volume sources offer the promise of great reliability and long lifetimes (>1 year), but have only been shown to work on test stands at very low duty factors (0.05 %) at 40 mA. Much development work is going into external antenna RF sources at SNS and CERN, only time will tell what performance is ultimately achievable.

Internal RF antenna volume sources have demonstrated high currents (>100 mA) at low duty factors (0.1 %) and 50 mA at 6 % duty factors, however in both cases lifetime is limited to a few weeks due to antenna wear.

Surface converter multicusp sources have demonstrated DC beams at reasonable $H^-$ beam currents of 20 mA but only with lifetimes of a few weeks and very long setup times (10 hours). Development sources have demonstrated 120 mA but with even shorter lifetimes.

Magnetrons have demonstrated high currents (80 mA) and very long lifetimes (>6 months) but only at very low duty factors (<0.5 %).

Penning sources offer high currents up to 60 mA (170 mA on experimental sources) and long duty cycles up to DC, however lifetime is limited to a few weeks.

Hot cathode multicusp volume sources have experimentally delivered DC currents up to 40 mA DC, and this current can be doubled with the addition of caesium. Filament lifetimes are short for high currents, but maintenance is very easy.

## 6.2    Power supplies

Ion sources present particular challenges for power supplies: they must deliver stable high currents to a plasma load that is often unstable. They must deliver stable high voltages to extraction electrodes that often breakdown. All of the electrodes that the power supplies are connected to are often in close proximity and often coated with caesium that increases the probability of sparking especially in an environment full of charge carriers in the presence of strong magnetic fields. All of these factors mean that the power supplies must not only be very robust and resistant to breakdown, but also very stable. Digital control electronics is susceptible to errors in the event of inevitable breakdowns, so analogue control circuitry is often a better choice.

## 6.3    Control systems

All of the ion sources discussed in this paper have to operate floating on a high-voltage platform, so it is essential to have some form of isolated control and measurement system to allow source tuning during operation and for the provision of timing signals. This is usually done over fibre optics, but some much older sources use insulated mechanical linkage to change power supply settings. The control system is invariably microprocessor based and must be very well isolated from all of the power supplies and housed in a well-screened chassis.

## 6.4    Developing a source

Most of the sources discussed in this paper are still being developed by labs all over the world. Beam currents are being increased, duty cycles extended, emittances reduced and reliabilities improved. Modern finite-element modelling and computational fluid dynamics allow the electrical, magnetic and thermal operation of the source to be investigated to a detail never before possible. Beam transport and plasma codes allow extraction and beam formation to be studied. All of these computer modelling tools allow ion source development to progress without having to go through as many prototype iterations.

It must be remembered, however, that ion sources and plasmas are incredibly complex: they exhibit numerous emergent behaviours that could never be predicted by simulation alone. The only way to find out how a new source design will behave is to actually test it. This is why development rigs or test stands are essential to designing new sources. These test rigs should replicate the actual environment where the source will run; ideally it should also include a LEBT to test exactly what beam can be transported.

A development test rig must be equipped with as many diagnostics as possible to try to understand how the source is performing.

These could include:
- beam current, e.g. toroids, Faraday cups;
- emittance, e.g. slit–grid, pepperpot, slit–slit, Alison electric sweep scanner;
- profile, e.g. scintillator, wire scanner, laser wire scanner;
- energy spread, e.g. retarding potential energy analyser;
- optical spectroscopy;
- Langmuir probes.

Sources must also run 24 hours a day if lifetimes are to be tested. Even then the true performance of source will not be known until it runs on an operational machine for several years, exposed to variable conditions and the inevitable human error.

## 7    Summary and conclusions

Meeting the beam current, pulse length and emittance required by the accelerator is only part of the job of an ion source. Operational sources must be reliable and they must have a lifetime that is compatible with the operating schedule of the accelerator. They must be easy to maintain so that in the event of a failure they can be easily fixed. If they have to be replaced they should be easy to dismantle. The start-up procedure should be made as easy and quick as possible.

Ion sources are a very interesting and stimulating area to work in. Sources cover a huge range of different disciplines, requiring skills in both engineering and physics. It can take a lifetime to become an expert in just one type of source. This paper has provided an introduction to some of the most common sources in use today in high-power hadron accelerators.

For further reading see the bibliography given in the Appendix. Huashun Zhang's book, *Ion Sources*, contains comprehensive references for every type of ion source discussed in this paper.

## Acknowledgements

I wish to thank Marthe Bacal, Jim Alessi, Martin Stockli, Rod Keller, Reinard Becker, Jacques Lettry, Raphael Gobin, Andrew Holmes, Joe Sherman, Alan Letchford, Scott Lawrie and Melisa Akdoğan for their comments and assistance during the writing of this paper.

## Appendix: Bibliography

### A.1. Papers

**A.2. Internet**

M.P. Stockli, *Ion Source 101*, Internet, 2001.

Wikipedia.

**A.3. Books**

Huashun Zhang, *Ion Sources* (Science Press, 1999).

Ian G Brown, *The Physics and Technology of Ion Sources* (Wiley-VCH, 2004).

Bernhard Wolf, *Handbook of Ion Sources* (CRC Press, 1995).